\documentclass{article}

\usepackage{arxiv}

\usepackage[utf8]{inputenc} 
\usepackage[T1]{fontenc}    
\usepackage{hyperref}       
\usepackage{url}            
\usepackage{booktabs}       
\usepackage{amsfonts}       
\usepackage{nicefrac}       
\usepackage{microtype}      
\usepackage{lipsum}		
\usepackage{graphicx}
\usepackage{natbib}
\usepackage{doi}

\usepackage{amsmath}       

\title{MoSES\_2PDF: A GIS-Compatible GPU-accelerated High-Performance Simulation Tool for Grain-Fluid Shallow Flows}

\date{} 					

\author{Chi-Jyun~Ko \\
        Dept. Hydraulic and Ocean Engineering\\
	National Chen Kung University\\
	70101 Tainan City, TAIWAN\\
        \And
	Po-Chih~Chen \\
        Dept. of Computer Science and Information Engineering\\
	National Chen Kung University\\
	70101 Tainan City, TAIWAN\\
        \And
	Hock-Kiet~Wong \\
        Dept. Hydraulic and Ocean Engineering\\
	National Chen Kung University\\
	70101 Tainan City, TAIWAN\\
        \And
        \href{https://orcid.org/0000-0001-5925-8207}{\includegraphics[scale=0.06]{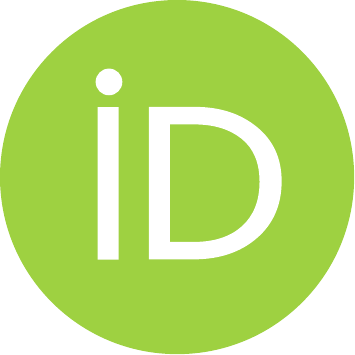}\hspace{1mm}Yih-Chin~Tai*}\\
	Dept. Hydraulic and Ocean Engineering\\
	National Chen Kung University\\
	70101 Tainan City, TAIWAN\\
	\texttt{yctai@ncku.edu.tw} \\
}

\renewcommand{\shorttitle}{A GIS-Compatible GPU-accelerated High-Performance Simulation Tool}

\hypersetup{
pdftitle={MoSES\_2PDF: A GIS-Compatible GPU-accelerated High-Performance Simulation Tool for Grain-Fluid Shallow Flows},
pdfauthor={C.J.~Ko, P.C.~Chen, H.K.~Wong and Y.C.~Tai},
pdfkeywords={GPU-accelerated, CUDA, GIS-compatible, Grain-fluid shallow flows, Simulatiom tool},
}

\begin{document}
\maketitle

\begin{abstract}
We introduce a GPU-accelerated simulation tool, named Modeling on Shallow Flows with 
Efficient Simulation for Two-Phase Debris Flows (MoSES\_2PDF), of which
the input and output data can be linked to the GIS system for engineering
application. MoSES\_2PDF is developed based on the CUDA structure, so that it
can well run with different NVIDIA GPU cards, once the CUDA vers. 9.2 (or higher) is installed.
The performance of the MoSES\_2PDF is evaluated, and it is found that the present 
GPU-CUDA implementation can enhance the efficiency by up to 230 folds, 
depending on the PC/workstations, models of GPU card and the mesh numbers in the computation domain.
Two numerical examples are illustrated with two distinct initial inflow conditions, which are included
in two modes of MoSES\_2PDF, respectively. In the numerical example of a large -scale event, the 2009 Hsiaolin event, 
the results computed by two distinct NVIDIA GPU cards (RTX-2080-Ti and Tesla-V100) 
are found to be identical but tiny deviation is figured
out in comparison with the results computed by the conventional single-core CPU-code. It is speculated to be caused by
the different structure in the source codes and some float/double operation.
In addition to the illustration in GIS system, the computed results by MoSES\_2PDF can also 
be shown with animated 3D graphics in the ANSI-Platform, where the user can interact with 3D scenes.
The feasibility, features and facilities  of MoSES\_2PDF are demonstrated with respect to the two numerical examples
concerning two real events.
\end{abstract}


\section{Introduction}
Many of the geophysical hazardous flows, such as landslides or debris flows,
usually take place in mountain area and flow on non-trivial topography. 
It is clear that the geometry of the topography may have significant impacts on the flow paths.
For the sake of hazard assessment,  risk management or evaluation of disaster mitigation countermeasures,
numerical simulation is commonly a powerful tool for scenario investigation.  
Hence, it is highly requested to develop an effective tool of high efficiency and high performance 
for providing useful information concerning the flow behavior, such as the paths, velocity and flow thickness, etc.
A successful simulation of hazardous flows on real topography depends on the availability of
the  digital elevation model (DEM) essential. With the rapidly growing popularity of the UAV and the integrated
modern remote sensing technology, the expensive high resolution data,
such as the LiDAR-derived DEM or the aerial photos, has become affordable and 
enriched the Geographic Information System (GIS) a lot.
From the viewpoint of scenario investigation, one of the urgent tasks would be the integration 
of the simulation tools with the GIS as well as with a high efficient
computational performance.
Numerical simulation based on Physics-based dynamical models can help the quantitative investigation on the impacts 
caused by the plausible geophysical hazardous flows. However, in addition to the high-resolution DEMs, 
it also requests expensive computer resources,
depending on the resolution and computation domain. The launch of Compute United Device Architecture (CUDA), 
a parallel computing platform and programming model, provides an econamic solution for speeding up the computational
efficiency by means of employing the graphical processing units (GPUs).
The computation duration can therefore be shorten to be less than the scenario time. Hence, either computation
with real-time  illustration or  training for model parameters has become achieveable in an efficient way
even without expensive computer resources.

In the past decades, many efforts have been devoted to modeling these highly destructive
flows. Since these flows are generally thin (measured normal to the basal surface) in comparison 
with the large extension in the direction tangential to the topographic surface, shallowness
assumption and depth-integration are often introduced for reducing the complexity of the model 
equations as well as of the numerical approach. For the ease of deriving the governing equations,
many of the models are given in Cartesian coordinates \citep[e.g.,][]{pitman2005two,li2011fully}, 
where the topography is added on the horizontal
or inclined plane, or in the {\em locally inclined Cartesian-type} 
coordinates~\citep[e.g.,][]{pudasaini2012general,pudasaini2019multi}.
For an appropriate description of the depth-averaged velocity as well as for the employment of the shallowness
assumption, the terrain-fitted coordinate system has been introduced. In addition to the curved-twisted
coordinate proposed by \cite{pudasaini2003rapid}, the remarkable works for general topography are
\cite{Bouchut2004,TaiKuo,tai2012alternative} and \cite{luca2016shallow}. As well,
assuming the non-trivial topographic surface to be composed of finite number of triangles,
local coordinate system is introduced on each triangular element
for reducing the complexity of numerical implementation in a unstructured mesh system 
\citep[see][]{rauter2018finite,tai2021multi}.


Regarding the composition of the flow body, models for hazardous flows can be divided into four categories: 
a) single-phase model; b) quasi-two-phase models; c) two-phase models; and d) multi-phase approaches.
The key issue by the single-phase approach is the introduction of an appropriate rheological relation, 
which is definitely different to the Newtonian one due to the complex composition. Taking into account 
the existence of yield stress, models with the Bingham plastic relation \citep[e.g.,][]{liu1989slow,coussot1997mudflow} 
or with the Herschel-Bulkley law \citep[e.g.,][]{huang1998herschel,ancey2007plasticity} 
have been suggested. In addition, a generalized visoplastic model is proposed by Chen (1988), 
covering the models by Bagnold\rq s model \citep{bagnold1954experiments} and Takahashi\rq s 
constitutive relation \citep{takahashi1978mechanical}. 
For the ease of numerical implementation, commercial codes are mainly released based 
on the single-phase models for engineering purposes. 
One of the most popular ones is the FLO-2D, initially following the rheological model 
proposed by \cite{o1985physical,o1988laboratory}
and improved by many sequential works \citep[e.g.][]{o1993two}.

The quasi-two-phase or the two-phase models are based on the concept of two-phase mixture. However,
in the quasi-two-phase models, the relative velocity between the two constituents is assumed to be very small, 
so that its impacts are assumed to be insignificant and only one velocity is considered for the
conservation of linear momentum \citep[e.g.,][]{iverson1997physics,iverson2001flow,pudasaini2005modelling,tai2012modelling}.
Since only one velocity is taken into account, the dynamic behavior of the flow body is similar to a single-phase 
fluid (if entrainment or deposit at the bottom is not included), and the effects of the concentration 
are then parametrically (instead of dynamically) 
considered in the rheological relation. In addition to the type of Coulomb-mixture~\citep{iverson1997physics,iverson2001flow}, 
another approach is the quasi-two-phase grain-flow model, widely employed in Japan for engineering applications,  e.g., 
\cite{egashira1997constitutive,egashira2007review,nakatani2008development,liu2013effect,nakatani2016case}.
Emphasizing the the role of the interstitial fluid on the behavior of flows, the Coulomb-mixture theory has been 
improved by introducing the non-hydrostatic pore-fluid pressure for the dilatancy effects by the solid constituent 
\citep{iverson2009elements,iverson2014depth}.

In the two-phase models, each constituent is described by individual momentum equation and has
its own velocity.  Hence, the key feature of a two-phase model is the capability of describing the phase separation, 
which is documented in field surveys and laboratory experiments. To our limited knowledge, the first inspiring two-phase model 
for debris flows would be \cite{pitman2005two}, although the fundamental investigation on two-phase mixture theory 
for grain-fluid has been made since years \citep[e.g.,][]{anderson1967fluid}. Based on \cite{pitman2005two}, 
\cite{pelanti2008roe} introduced the well-balancing property in the numerical implementation, 
where  the variable topography is included. \cite{pailha2009two} suggested a two-phase description 
for the initiation of underwater granular avalanches.
\cite{luca2012modeling} has theoretically considered over-saturated mixtures in coordinate of general topography.
The quadratic viscous drag force and virtual mass force are considered in \cite{pudasaini2012general}.
A two-phase model in two-layer type and taking into account the dilatancy effects can be found in \cite{bouchut2016two},
which mainly considers the submarine debris flows, i.e. the upper layer is pure fluid and not thin.
With two-layer approach, the partially unsaturated grain-fluid mixtures are modeled by \cite{meng2017modeling},
where a fluid-saturated granular layer is overlaid by a pure granular material. 
With thermodynamically consistent derivation, models of granular-fluid mixtures with non-hydrostatic pore pressure can be found in 
\cite{hess2017thermodynamically,hess2019debris}, where the solid phase is assumed hypoplastic.
The multi-phase approach can be found in \cite{pudasaini2019multi}, where the flow body is 
composed of three phases (the coarse and fine solid fractions, and the viscous interstitial fluid),
and the model equations are given in the {\sl locally inclined Cartesian-type} coordinates for
describing the flows over a non-trivial topography.

Regarding the GIS-compatible simulation tool, \cite{mergili2017r} introduced a GIS-supported open-source 
computational framework, r.avaflow,
which is developed to link to the open-source GIS package, GRASS GIS.
The r.avaflow is based on a two-phase solid-fluid mixture model \citep{pudasaini2012general},
and mass exchange at the basal surface (entrainment or deposition) are taken into account.
In addition, r.avaflow allows multiple-model-runs that multiple parameter sets can be simultaneously executed
with the parallel processing function. Although it does not shorten the duration for a single run (a parameter set), 
this facility may help in looking for the optimal parameter set or collecting abundant plausible
simulation scenarios. The r.avaflow has been extended for the multi-phase 
flow model \citep{pudasaini2019multi}, see~\cite{webpageavaflow}, and several
engineering applications are available in \cite{mergili2020back,baggio2021advances}.


Although many of the above-mentioned models are with numerical implementation
and able to provide satisfactory results, 
the computation processes are usually very time-consuming and cost a lot of computer resources.
In order to improve the efficiency of computation, 
one may speed up the computing applications by harnessing the power of 
graphical processing units (GPUs).
Although GPU was originally designed for graphics rendering on computers, especial for 
the illustration of 3D objects, GPU-accelerated
high-performance computation has become more and more popular in various scientific
fields, such as Artificial Intelligence (AI), machine learning (ML), deep learning (DL) etc. 
For example, \cite{castro2011gpu},  \cite{brodtkorb2012efficient}, \cite{de2013efficient},
and \cite{aureli2020gpu}
have proposed GPU-accelerated approach based on the shallow water equations (SWEs), while
\cite{dazzi2019integration} integrated the GPU-accelerated 2D SWEs code with levee breach 
mechanism. A GPU-enhanced 2D SWEs code for curved triangular meshes is developed by
\cite{wu2021high}, of which the scheme is entropy stable and  provides high-order accuracy.

However, there is no GIS-compatible and GPU-accelerated simulation tool for two-phase grain-fluid flows yet.
Based on the two-phase model by \citet{tai2019modeling}, 
we developed the GPU-accelerated simulation tool, MoSES\_2PDF, which follows
the CUDA structure to integrate the GPGPU technology.
In addition, the MoSES\_2PDF is designed to be GIS-compatible.
That is, the input data is in the asc-format and
can be directly acquired from GIS systems (ArcGIS, QGIS), and
the output results can be transformed into asc-format for illustration in GIS
systems.
On the other hand, in MoSES\_2PDF we also use the GPU graphics
rendering functions in OpenGL for 3D numerical results presentation,
which combines the computation results with 3D objects to mimic
the effect of reality simulation. 

In the second section, we brief the grain-fluid model and employed
numerical implementation. The introduction of the framework of MoSES\_2PDF
and its performance with the GPU parallel computation facility 
are given in Sect.~\ref{Sect3}, two modes for different initial inflow conditions
are designed and three output formats available for results illustration. 
In the fourth section, two application examples are illustrated, which are
two historical events taking place in 2009 in Taiwan.
The application potential and the key features of
MoSES\_2PDF are summarized and highlighted in the concluding remarks.

%
%
\section{Modeling and Numerical Implementation}
\label{Sect2}



\subsection{Modeling}
\begin{figure}[t]
   \centering
\includegraphics[width=11cm]{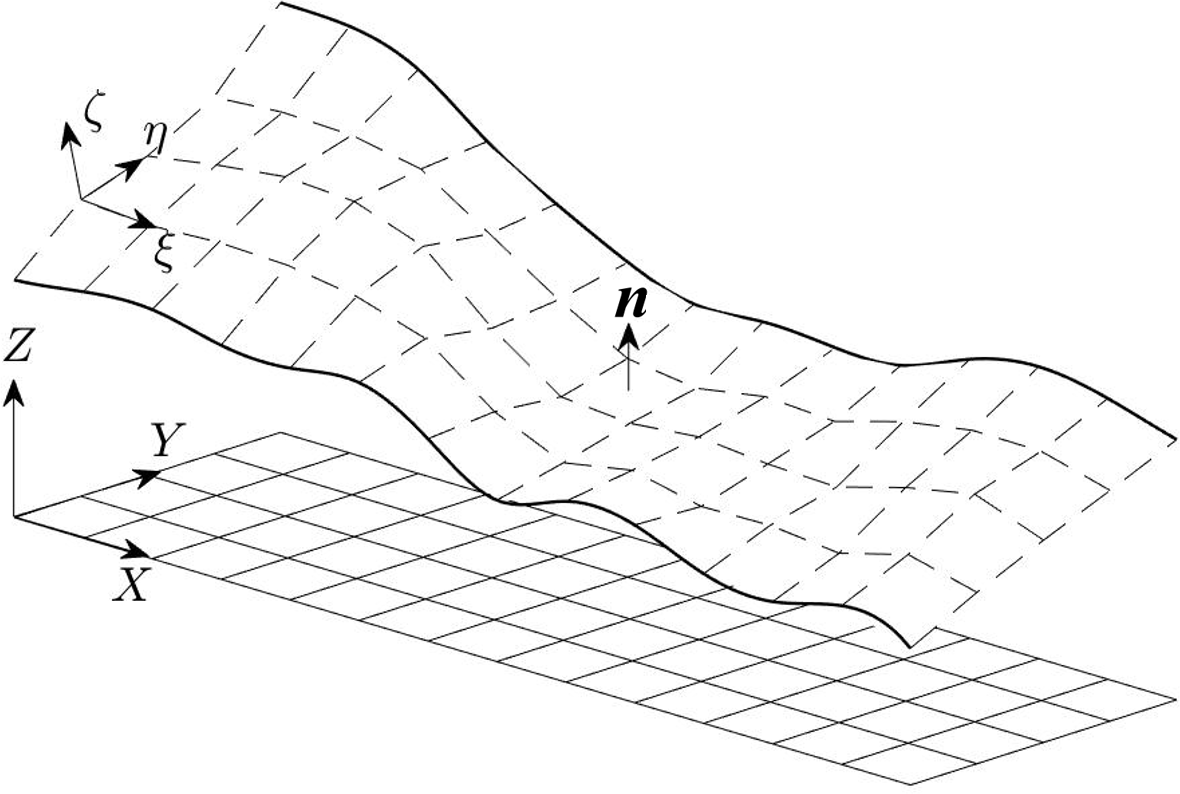}
\caption{Coordinate system $O_{XYZ}$ and ${O}_{\xi\eta\zeta}$.}
\label{Coord}
\end{figure}

In the GIS system, the topography is described by a set of altitudes of terrain locations
over a regular horizontal grid, the digital elevation model (DEM). Based on the DEM, a terrain-fitted coordinate system
${O}_{\xi\eta\zeta}$ can be introduced in the vertical-horizontal-oriented 
Cartesian coordinates ${O}_{XYZ}$, where the $X$- and $Y$-axis lie on the horizontal
plane and the direction of $Z$ points upwards. 
The system ${O}_{\xi\eta\zeta}$ can be chosen in the way
that, e.g., the projections of $\xi$-axis
and $\eta$-axis on the horizontal plane coincide with $X$-axis and $Y$-axis, respectively 
(cf. Fig.~\ref{Coord}). The coordinates between $O_{XYZ}$ and
 $O_{\xi\eta\zeta}$ is related by the transformation matrix ${\boldsymbol\Omega}$, 
 see~\citet{tai2012alternative} or~\citet{luca2016shallow} for details.
 
In MoSES\_2PDF, the evolution of the flow body is computed by a 
depth-integrated two-phase grain-fluid model for debris flows over rugged 
topography~\citep{tai2019modeling}, where  the conservations of mass and momentum 
are taken into account and both of the coordinate systems,
${O}_{XYZ}$ and ${O}_{\xi\eta\zeta}$, are used.
Hence, there are 6 model equations for the evolutions of the 6 physical 
quantities, $(h^{s,f}, v_X^{s,f}, v_Y^{s,f})$. Here and hereafter,
the superscripts \lq\lq $s$\rq\rq\ and \lq\lq $f$ \rq\rq\ indicate
the variables for the solid and fluid constituent, respectively, and the subscript $_{X,Y,Z}$
indicates component in $O_{XYZ}$, while $_{\xi, \eta, \zeta}$ for the one in $O_{\xi\eta\zeta}$.
For a concise description, the employed model equations are not shown here, but presented
in \ref{AppendixA} for readers. 

Since the flow body is treated as a two-phase mixture, 
in $(h^{s,f}, v_X^{s,f}, v_Y^{s,f})$,  $h^s=h\phi^s$ and $h^f=h\phi^f$ are defined, 
where $\phi^{s,f}$ 
 denoting the depth-averaged volume  fraction and
$h$ represents the flow thickness measured in the direction normal
 to the basal topography.
In addition, it should be noted that the variables $v_X^{s,f}$ and $v_Y^{s,f}$
are the $X$- and $Y$-component of the flow velocity projected on the horizontal plane.
With the help of the transformation matrix ${\boldsymbol\Omega}_b$ on the basal surface, 
one can get the flow velocities tangential to the topographic surface.

%
%
\subsection{Numerical implementation}
\label{Sect_Num}
MoSES\_2PDF is developed based on the CUDA structure  (CUDA Toolkit vers. 11.0). 
The employed numerical implementation is same as in \citet{tai2019modeling},  that
the balance equations with respect to the solid and fluid constituent are computed separately
at each time step, where the anti-diffusive, non-oscillatory central scheme 
(adNOC, see~\cite{kurganov2000new, kurganov2007second}) 
is used for solving the six differential equations.
For details of the numerical implementation, we refer the readers to \citet{tai2019modeling},
in which the code is developed for single-processor (single-core) computation.

The high resolution in space is achieved by
utilizing the cell reconstruction, such as the total variation diminishing (TVD), 
essentially non-oscillatory (ENO) or weighted essentially non-oscillatory (WENO) scheme. 
And the high-order accuracy in time is carried out by the multi-step Runge-Kutta (RK) method
with respect to a semi-discrete formulation given in \cite{kurganov2000new}. 
In MoSES\_2PDF, the Minmod TVD slope limiter is adopted and 
the two-step modified Euler (second-order RK) is employed. As indicated in \cite{kurganov2000new}, the time 
step should be limited by the Courant-Friedrichs-Lewy (CFL) condition with a number of 0.125 and 
the CFL = 0.1 is set in MoSES\_2PDF.

%




%
%
\section{Framework of MoSES\_2PDF and its Performance}
\label{Sect3}

%
%
\subsection{Framework}

\begin{figure*}[t]
  \centering
\includegraphics[width=13.2cm]{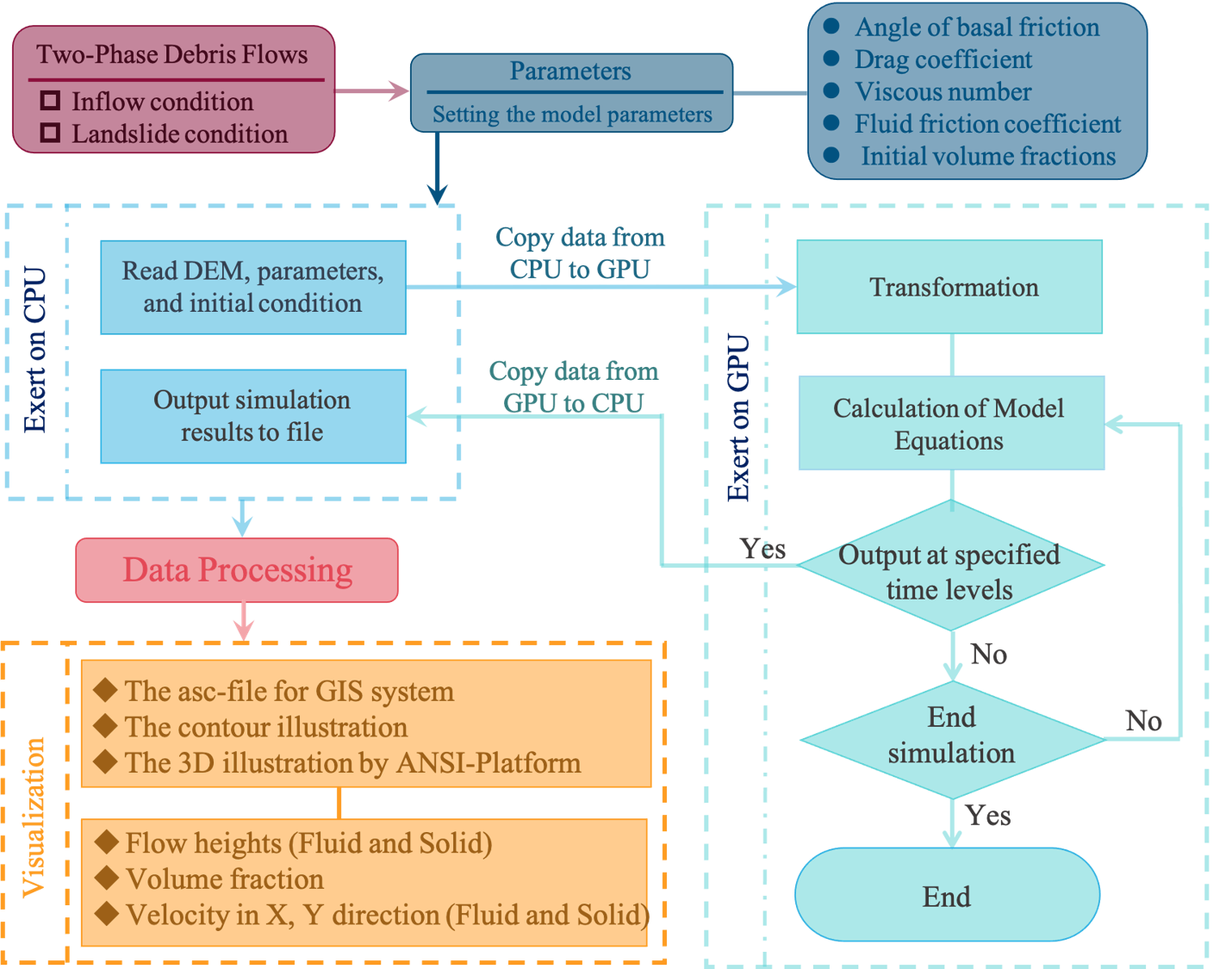}
\caption{The framework of computation}
\label{FlowChart}
\end{figure*}

%
%
\subsubsection{Teamwork of CPU and GPU}
\label{GPU_GPU}

With respect to the hardware structure, the computational 
framework of MoSES\_2PDF can be
divided into two sections, the CPU- and GPU-section, where they have to
communicate with each other to complete the operation of computation. 
Since the data transfer between the sections could be very
time-consuming, it is therefore designed in the way to minimize
 the data transfer as possible, that the CPU-section mainly processes file
reading and output, and the rest computational procedures are
operated in the GPU-section, cf. Fig.~\ref{FlowChart}. 

According to the characteristics of applied numerical scheme,
the execution hierarchy of single instruction corresponding 
to multiple threads (SIMT)  is employed as the GPU
computing architecture for the purpose of a high efficiency in computation. 
Besides, most of the calculations are matrix operations, 
yielding the fact that
the memory location of a matrix may cover several blocks, 
depending on the size of the computational domain (grid numbers).
It leads to a hurdle in looking for the global extreme (maximum or minimum) value in the matrices,
because the function of global synchronization is not supported in the 
CUDA parallel-programming model (library).
If one uses the type of loop for searching over the whole domain in the CPU-section, 
the data transfer/collection among the blocks
will significantly retard the computational efficiency.  
Hence, the extreme-value-searching plays a crucial role in efficiency enhancement, 
and one has to find an effective way for data communication among different blocks.
In MoSES\_2PDF, we adopt the method of optimizing
parallel reduction, proposed by~\cite{harris2007optimizing}, in the CUDA code to compute
the maximum/minimum values in the entire domain.  That is, taking the maximum value as example,
the maximum value on each block is first figured out, and then 
they will be transferred to a specified block, in which the global maximum value is identified. 
In this way, the computation efficiency of the searching process is  improved with a significantly higher efficiency than
the computation of using the loop type.

%
%
\subsubsection{Operation of MoSES\_2PDF}

The operation of MoSES\_2PDF consist of  four main stages (cf.~Fig.~\ref{FlowChart}):
(0) Preparation;
(1) Input and parameter setting; (2) Computation; and (3) Results illustration:\\[0.2cm]
\noindent{\bf Stage 0:} Preparation \\[0.2cm]
Based on different initial inflow types at the boundaries of computational domain,
MoSES\_2PDF provides two modes: 
In the first mode (Mode-I), finite masses with specified concentration(s) 
are  to be released in the computational domain, where there is no mass inflow from the boundary
in the initial condition. This mode is mainly used for the simulation of landslide type,
and this type of releasing  finite mass(es) can be found in many works 
\citep[e.g.,][]{Gray1999,pitman2005two,Kuo2009,mergili2017r,pudasaini2019multi,de2019imex_sflow2d}.
The second mode (Mode-II) is used when there are materials flowing into the computational
domain through the boundaries. That is,  the inflow condition
(the hydrograph of the flow thickness, volume fraction (concentration) of solid/fluid phase,
velocities and the inflow locations) at the specified locations of boundary can be defined in Mode-II.
This kind of inflow condition is often adopted in debris flow simulations, such as the debris flow simulator
Kanako 2D~\citep[e.g.,][]{nakatani2008development,liu2013effect,nakatani2016case}, developed by the
team in Kyoto university, Japan.  
In both modes, no reflection boundary condition is set in the whole process of computation.
In addition to choosing the suitable mode, one has to prepare the DEM of the specified computational domain 
and its associate initial condition (the flow thickness, concentration of solid/fluid phase and
velocity distribution of the flow body). The DEM is in the asc-format, which can be directly extracted 
from GIS system (ArcGIS or QGIS). According to the initial inflow type, 
the data file, the initial configuration for Mode-I or the inflow condition for Mode-II, should be prepared 
(in ASCII-format).
\\[0.2cm]
\noindent{\bf Stage 1:} Input and parameter setting \\[0.2cm]
%
Once the code is initiated, a text-file \lq\lq par\_list\rq\rq\ is to be read first,
on which all the model parameters, the total
simulate time, the time interval for results output, 
and the locations of the DEM file, inflow/initial condition file
can be specified. The model parameters consist of: 
(a) the angle of basic friction $\delta_{b}$ between the solid 
constituent and the basal surface;
(b) the drag coefficient $C_{d}$ between the solid and the fluid constituents, 
caused by the related velocity between the two constituents;
(c) the viscosity coefficient $N_{R}$,  similar to the Reynolds number with value 
inversely proportional to the viscosity; and
(d) the fluid friction coefficient $\vartheta_b$. 
Since the employed model is for two-phase solid-fluid flows, the initial 
concentration of solid phase 
 $\phi^{s}_{0}$ is also given in the text-file  \lq\lq par\_list\rq\rq.\\[0.2cm]
\noindent{\bf Stage 2:} Computation \\[0.2cm]
Once the reading process is completed in the CPU-section, 
all the data will be transmitted to the GPU-section for computing
the evolutions of the variables.
When it arrives at the specified time level for results output, the data
is transferred from the GPU memory to the host memory for the output
operation by CPU, cf.~Fig.~\ref{FlowChart}.

As elaborated in Sect.~\ref{Sect_Num}, the time interval should be determined
according to the assigned CFL condition, for 
which the global maximum wave speed is needed.
Since the time interval has to be determined at each time step, this process 
is the most time-consuming one.
With the method of optimizing
parallel reduction mentioned in Sect.~\ref{GPU_GPU}, the
efficiency of the model computation can be enhanced by ca. 20-fold.\\[0.2cm]
%
%
\noindent{\bf Stage 3:} Results illustration\\[0.2cm] 
The output of the computed results includes the total flow
thickness (measured normal to the basal surface), the fraction of solid phase, the $X$-
and $Y$-components of velocity for the solid and fluid phases, respectively.  
In the data processing, the results can be converted in three different types,
depending on the requirement of illustration platform.
The first is the asc-file format for illustration in GIS system 
(ArcGIS or QGIS) as well as for the
subsequent applications.
The second is in the format for drawing contour maps by Python,
usually for scientific investigation.
The third type is for a three-dimensional scenario illustration tool
(Advanced Numerical Scenario Illustration Platform, ANSI-Platform), developed
by Y.C. Tai and his coworkers with OpenGL. The ANSI-platform provides
animated 3D graphics with user-interactive facility, that
one can interact with 3D scenes (zoom in/out, translate or rotate) 
for examining the simulation results.
In case the corresponding ortho-photo of satellite image is available, 
the animated 3D topography can be integrated with the photograph on its surface
for the purpose of or scenario investigation, hazard assessment or 
disaster mitigation.


%
%
\subsection{Performance of GPU parallel computation}

The performance of the MoSES\_2PDF with/without GPU-CUDA computation
is evaluated by a campaign of the accumulated elapsed time (duration of computation) for running 10,000 steps 
with various number of meshes. In this campaign, computations by
single-process of PC (i7-8700 CPU@3.20 GHz, 16 GB memory, Linux OS) 
and by workstations with two different  models of NVIDIA GPU cards (Geforce RTX 2080 Ti 
with i7-8700 CPU@3.20 GHz and Tesla V100-SXM2-32GB with Xeon Gold 6154 CPU@3.00GHz)
are performed. Altogether, there are 6 numbers of meshes (i.e., 10,000, 50,000, 100,000, 250,000, 500,000 and 1,000,000) 
in the computational domain for evaluating the performance.

%
%

\begin{figure}[t]
  \centering
\includegraphics[width=11.2cm]{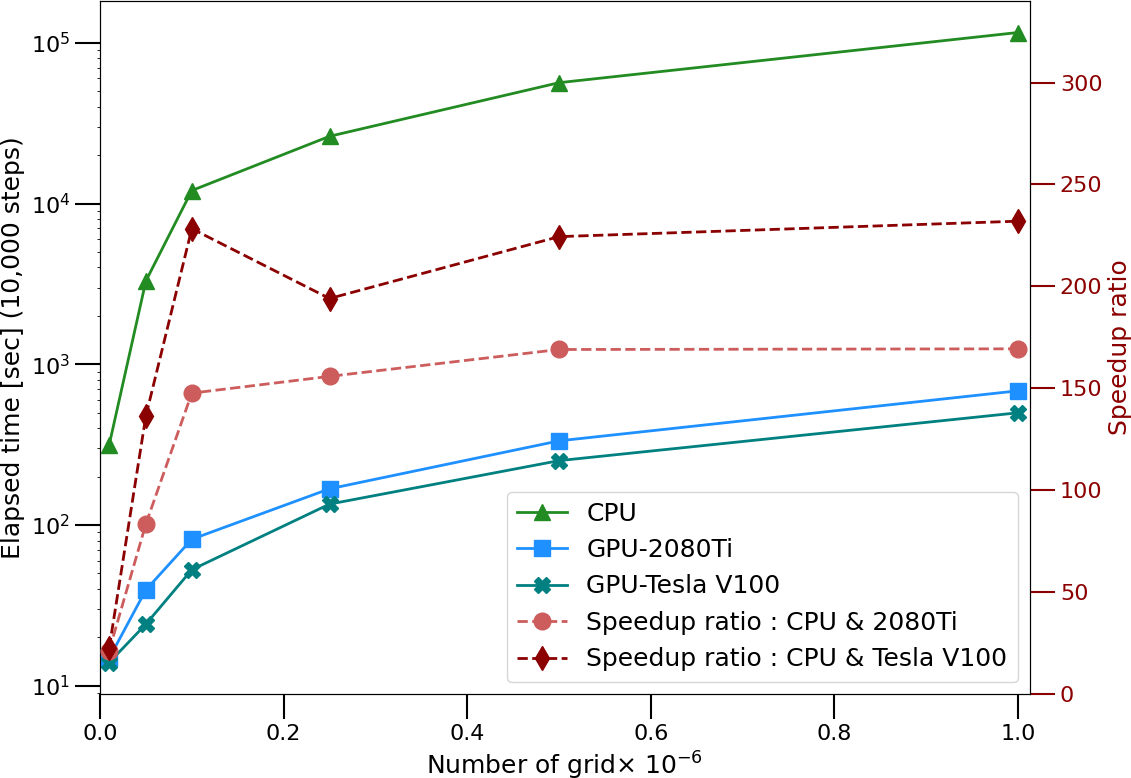}
\caption{Performance of CPU and GPU and the associated speedup ratio}
\label{CPU_GPU}
\end{figure}

The accumulated elapsed time is calculated by the averaged duration of three repeated 
runs for minimizing the time deviations by each computation.
As shown in Fig.~\ref{CPU_GPU}, the duration of computation increases for more meshes, where
the solid lines indicate the accumulated elapsed time and the dashed lines stand for the speedup ratio.
The speedup ratio reflects the increasing rate, of which the value is determined by
the duration for the GPU computation divided by the elapsed time for the 
CPU single-processor computation.
For computation with 10,000 meshes, the GPU speedup ratio is
about 20-fold, while its magnitude rapidly increases as the number of meshes grows up to 100,000. 
When the number of meshes reaches 500,000,
the speedup ratio exceeds 150-fold, while it is about 160-fold for computation with 1,000,000 meshes. 

It is interesting to find that there is one nick of speedup ratio at mesh number of  250,000
with respect to the computation by GPU-Tesla V100, although the elapsed time monotonically 
increases as the mesh number grows. This nick is suspected to be caused by the arrangement of memory allocation
in the Tesla-GPU card, e.g., the block size in threads as elaborated in~\cite{ryoo2007program}. 
In general, the enhancement of the efficiency is extremely significant, and the more the meshes
there are, the better the efficiency is improved. The employment of the GPU computation sheds the
light on the achievement of a real-time computation and illustration for scenario investigation of hazard assessment
or disaster management.





%
%
\section{Application examples}\label{Sect4}

Two application examples are introduced: the 2009 Hsiaolin landslide 
and the DF004 debris flow event. Both of the events  took place 
during the typhoon Morakot in 2009 and locate  at southern Taiwan. 
More than 470 people were killed in the 2009 Hsiaolin event, and 41 victims are identified during the debris flow
event DF004 \citep[cf.][]{dong2011,FastReport2009}.
In the 2009 Hsiaolin landslide event, all
the blocks of released mass lay in the computational domain
(see Fig.~\ref{HL_Fig2}a).
In the computation, the blocks are released simultaneously, sliding downward and attacking
the Hsiaolin village, so that this event is categorized as the landslide type and the MoSES\_2PDF-I 
(Mode-I) is utilized. In the DF004 debris flow event, a large amount of soil, sand and water rushed from upstream
into the Nansalu village along the gully, so that the Mode-II is used for the simulation.

In the two application examples, there are totally five parameters to be set, of which 
the four, $(\delta_b, C_d, N_R, \vartheta_b)$ ,
are the material parameters and $\phi^{s}_{0}$ is the solid concentration of the initial/inflow flow body.
For showing the feasibility of MoSES\_2PDF,
they are set identical to the ones used in previous studies~\citep{tai2019modeling,tai2020idealized}, 
as listed in Table~\ref{Tab_parameter}.
In addition, for ease of depicting the dynamic evolution of the flow body as well as identifying the flow paths,
only the area covered by flow thickness more than 10 cm will be shown in the results illustration, 
if not additionally specified.

%
%
\begin{table}[t]
  \caption{Parameter values used in the application examples}
  \centering
\begin{tabular}{c | c c c c c }
Parameter & $\delta_{b}$  & $C_{d}$ & $N_{R}$ & $\vartheta_b$ & $\phi^{s}_{0}$ \\
\midrule
   Value     &  $16^{\circ}$  & $6.0$     & $268$     &     $5.0$          & $0.5$\\
\end{tabular}
\label{Tab_parameter}
\end{table}

%
%
\subsection{The 2009 Hsiaolin Landslide}

The 2009 Hsiaolin Landslide is of an abnormal large-scale, that the largest scar
area caused by the released mass covers about 57 hectares and the total volume
exceeds 20 million cubic meters~\citep{dong2011,kuo2011landslide}.
Because of the huge size and the severe casualties,
many post-event investigations have been conducted.
Based on the difference of the pre- and post-event DEMs,  landslide blocks
are identified for reconstructing the flow paths, \citep[see][]{kuo2011landslide,tai2019modeling}.
The computational domain and mesh size are identical as in \citet{tai2019modeling}, 
that it covers $3.7\times 2.2$ km$^2$ and with resolution $\Delta x =  \Delta y = 10$ m.
Hence there are totally 85,956 cells, including 3 ghost cells at each boundary, in the computation.
For the sake of validating the GPU implementation in MoSES\_2PDF, identical parameters are adopted 
(see Table~\ref{Tab_parameter}), and the total simulation time is set to 181.82 s, when
the whole flow body is nearly at the state of rest.

With identical initial condition and parameters, the single-processor (core) 
computation and GPU parallel computation
with RTX 2080 Ti and  Tesla V100 are performed. The single-core 
computation took about 2.5 hours, while the GPU parallel computations
with  CUDA implementation took only ca. 80 s and 56 s with 
RTX-2080-Ti and Tesla-V100, respectively.
Although the GPU computations are identical (12,427 steps), 
they are slightly deviated from the CPU results (12,454 steps).
After a tedious checking process between the CPU- and GPU-codes as well as results, it is found that 
the difference of the first time step occurs at the eighth decimal place. The discrepancy is suspected to be
caused by the coding structure and cut-off error in the float/double operation in the GPU-computations. Nevertheless, 
regarding the final deposition area the difference is rather minor. At t = 181.82 s, the deposited mass covers 
14,872 meshes where only 25 are in difference, if the meshes with thickness less than 1 cm are isolated.
In case only flow thickness more than 10 cm is taken into account, there are 23 in 13,600 meshes to be deviated.
In both cases, the deviation is less than $0.17\%$. 
Hence, only the facility of GPU computation is provided in MoSES\_2PDF.

\begin{figure}[!ht]
  \centering
  \hbox{\hspace{2.5cm}a)}
  \includegraphics[width=8.8cm]{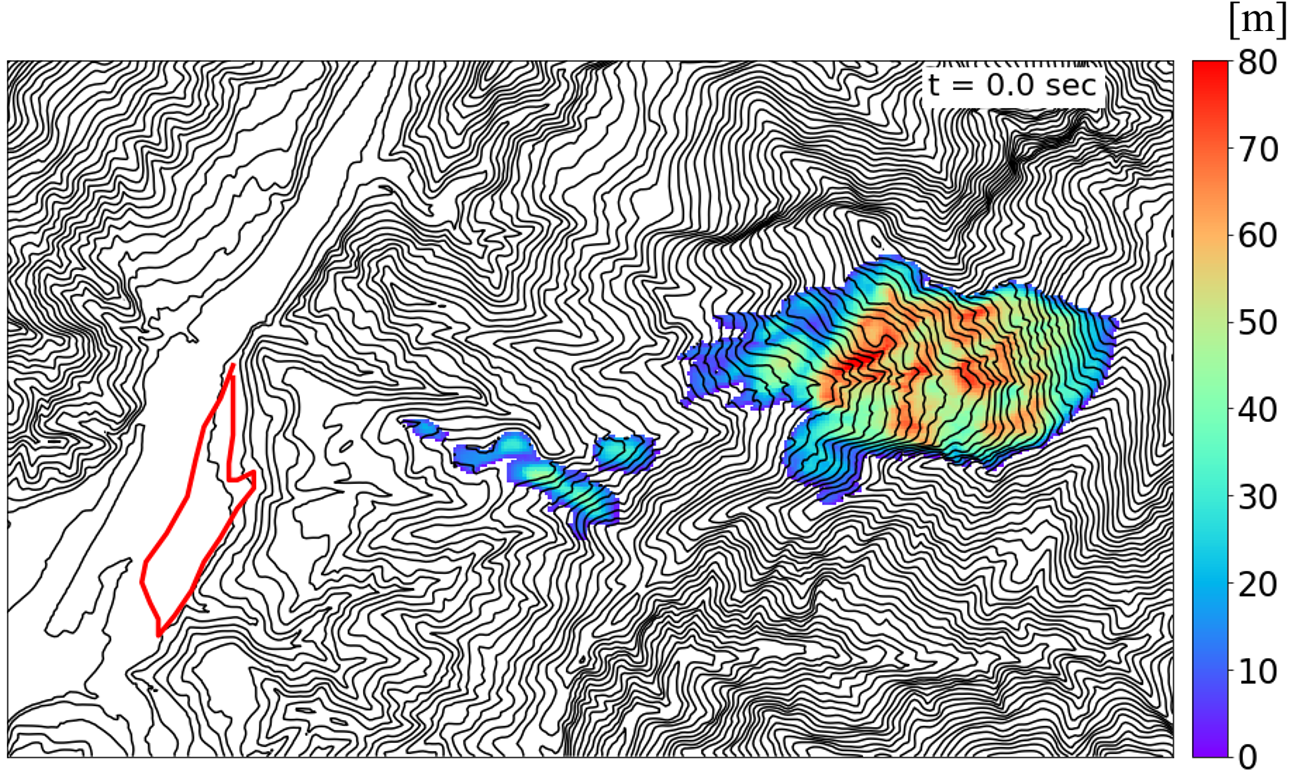}\\
  \hbox{\hspace{2.5cm}b)}
  \includegraphics[width=8.8cm]{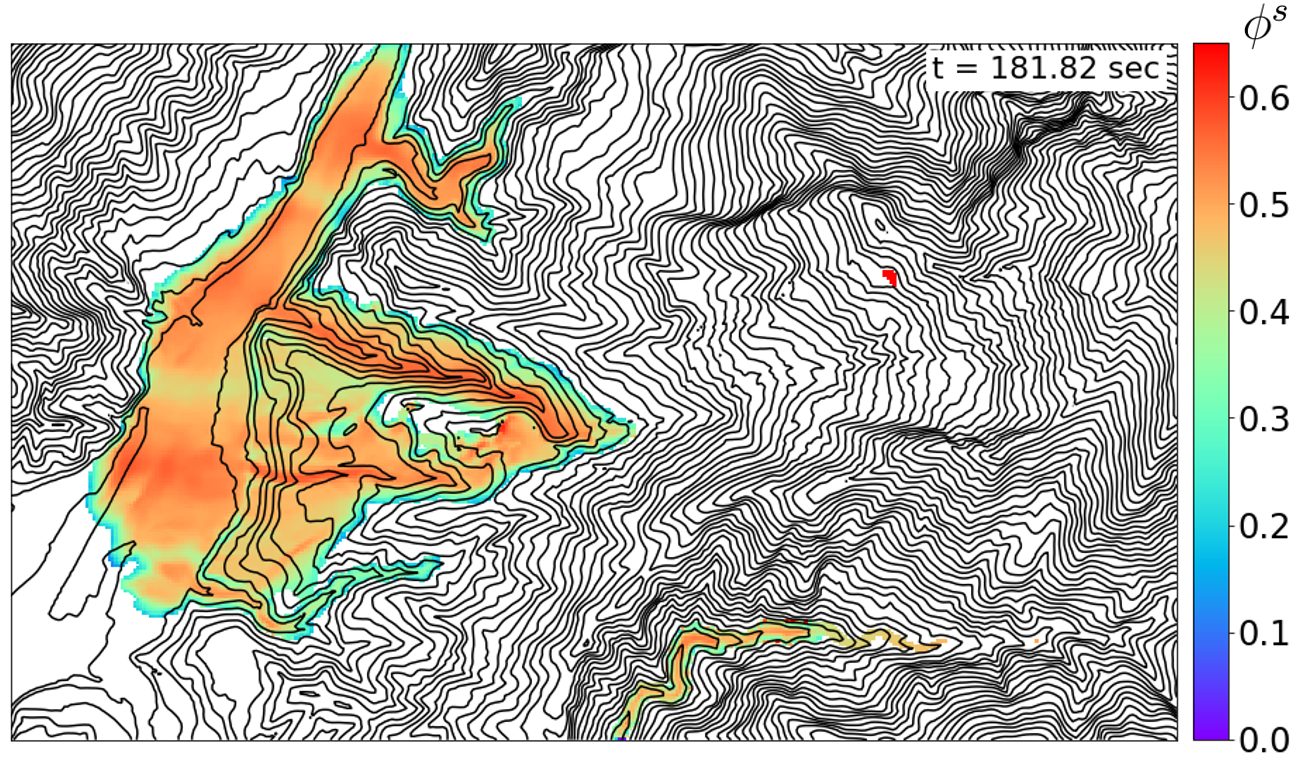}\\
  \hbox{\hspace{2.5cm}c)}
  \includegraphics[width=8.4cm]{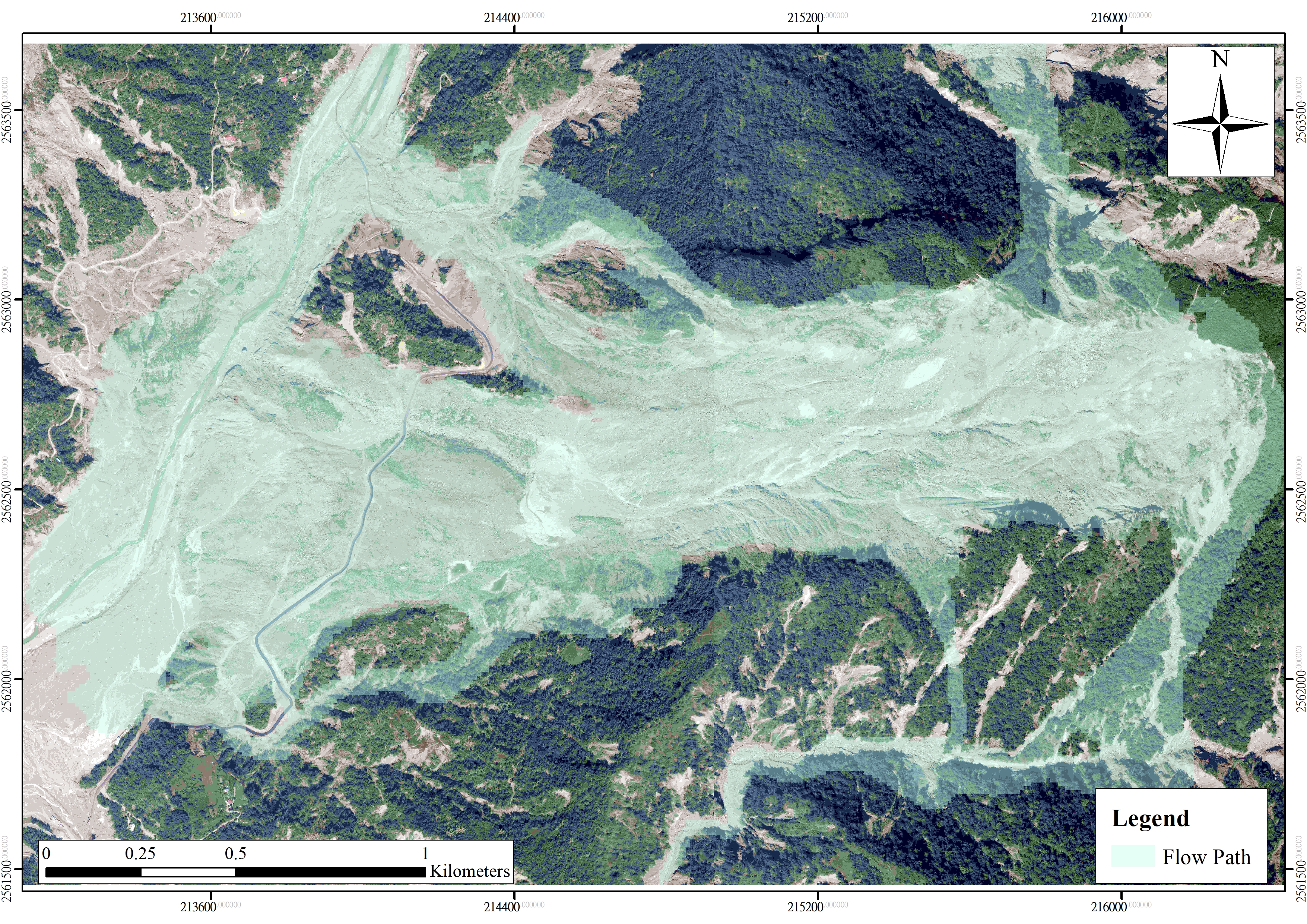}
\caption{Simulation of Hsiaolin event: (a) The initial flow thickness, where the red line outlines the Hsiaolin village; 
(b) Distribution of the solid concentration at $t = 181.83$ s; and (c) The computed flow path with the satellite orthophoto}
\label{HL_Fig2}
\end{figure}

Figure~\ref{HL_Fig2}b depicts the computed distribution 
of solid phase over the contour topographic map at $t=181.82$ s.  It is found that the 
high concentration of solid phase approximately coincides with the locations of valley.
In Fig.~\ref{HL_Fig2}c, the flow paths are indicated by the cyan shade, which is illustrated on the orthophoto
of satellite image taken about 6 months after the event. The bar areas in the satellite image
can be recognized beneath the cyan shade and they approximately point out the zones 
rushed by the landslide mass. The sound agreement reveals the feasibility of the MoSES\_2PDF 
with CUDA implementation for scenario simulation in high efficiency. 

\begin{figure}[t]
  \centering
\includegraphics[width=11.6cm]{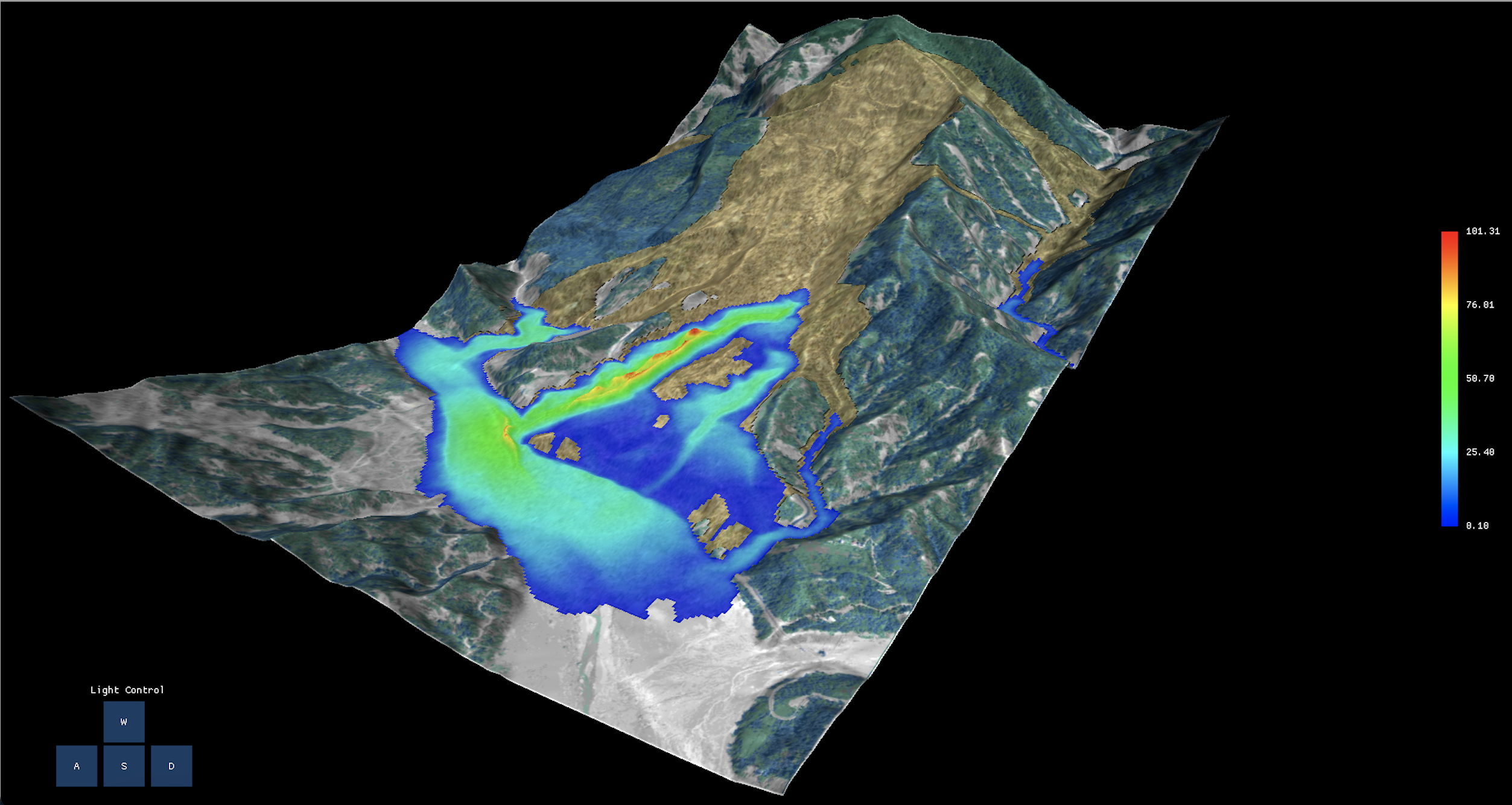}
\caption{The 3D terrain illustration in the ANSI-Platform: Simulation of the 2009 Hsiaolin event, where
the flow thickness is shown by the color map and the brown zones indicates the flow paths.}
\label{3D_Fig}
\end{figure}

Figure~\ref{HL_Fig2}ab are plotted by the illustration 
facility written by Python, and Fig.~\ref{HL_Fig2}c is constructed by the toolbox embedded in the QGIS.
Figure~\ref{3D_Fig} is one snapshot of the animated 3D scenes  in the ANSI-Platform,
where the surface of the topography is integrated with the satellite image.
In the ANSI-Platform, the dynamic distribution of flow body is depicted with color map, 
the accumulated flow paths are given by brown shade, where the partly transparent shade 
enables the structure or topographic features in the photographs to be recognized.

\subsection{The channel-type debris flow: DF004 event}

\begin{figure}[t]
  \centering
\includegraphics[width=14.2cm]{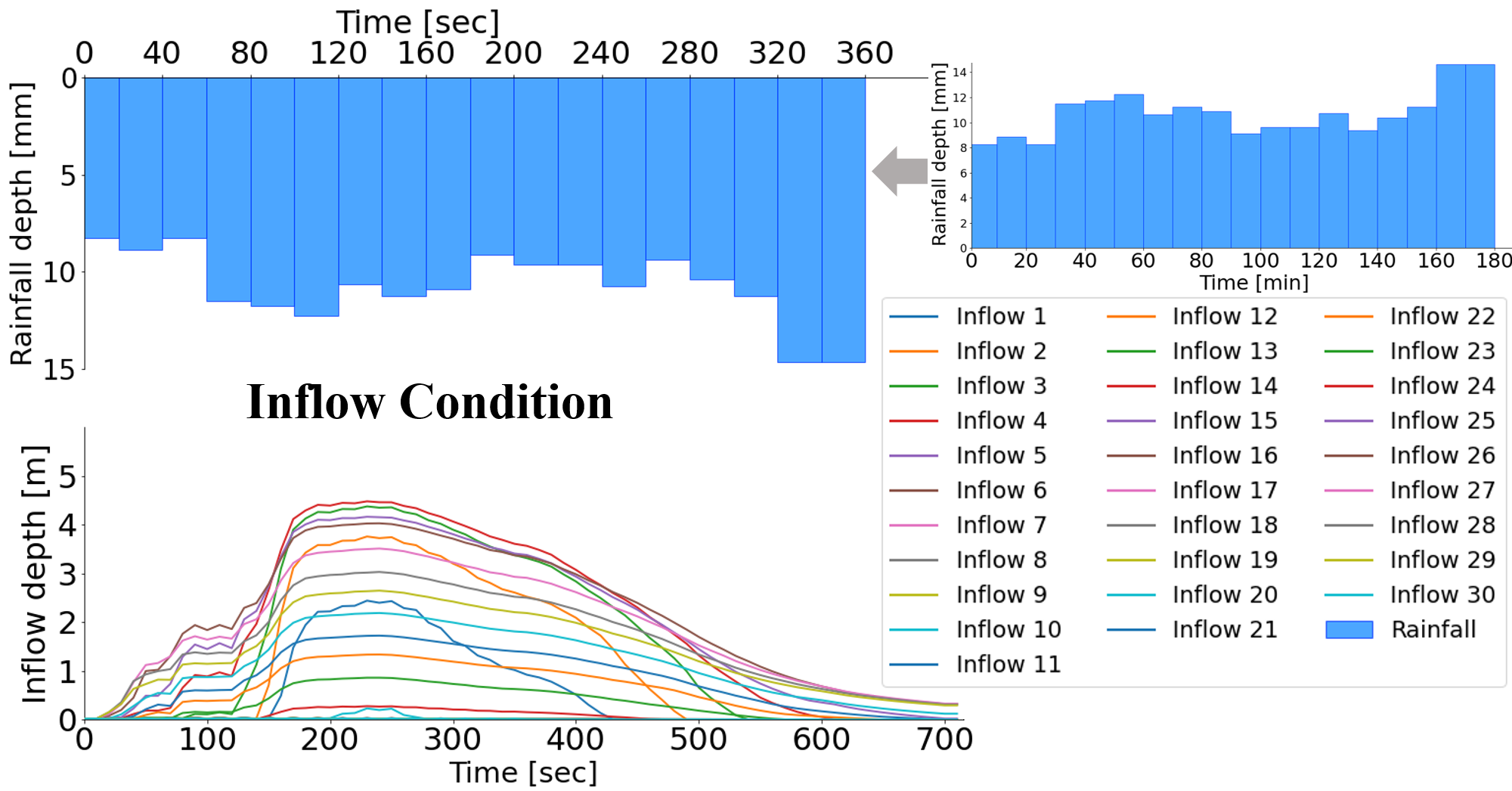}
\caption{The applied inflow condition for DF004 debris flow event}
\label{RainFallCondition}
\end{figure}

%
%

In the DF004 event, the debris flow rushed through 
the channelized valley and attacked
the Nansalu village after a long traveling distance around 17:00 on Aug. 9th, 2009. 
Large amount of debris 
were entrained from the river bed during the movement.
For isolating the complex process of collecting the debris during the traveling 
stage and improving the simulation efficiency, the computational domain focuses on the
encountered village, where appropriate inflow conditions are estimated 
based on the records of the nearest rainfall station several hours before 
the occurrence of the debris flow.

\begin{figure}[t]
  \centering
  \hbox{\hspace{2.5cm}a)}
  \includegraphics[width=8cm]{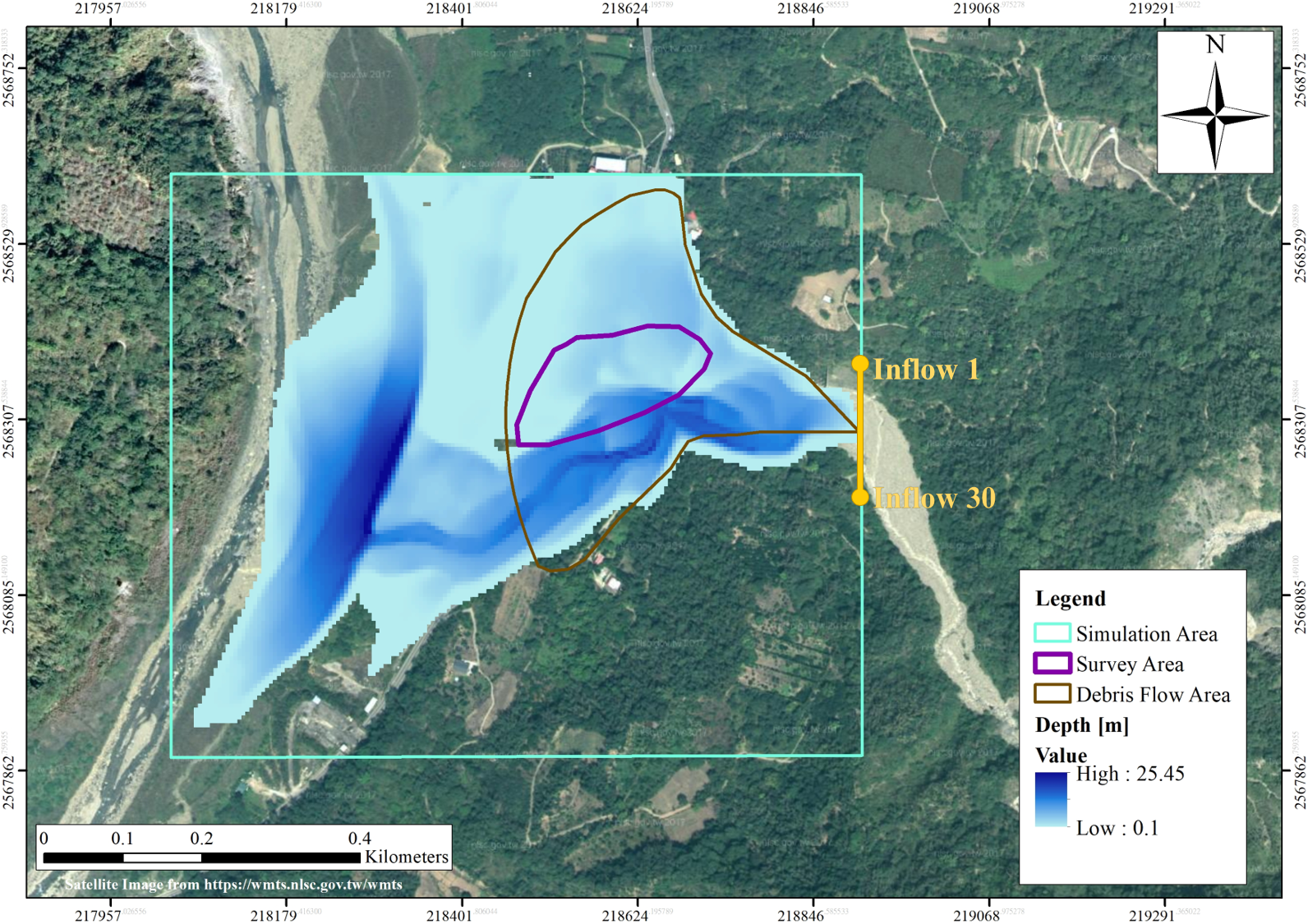}\\
  \hbox{\hspace{2.5cm}b)}
  \includegraphics[width=8cm]{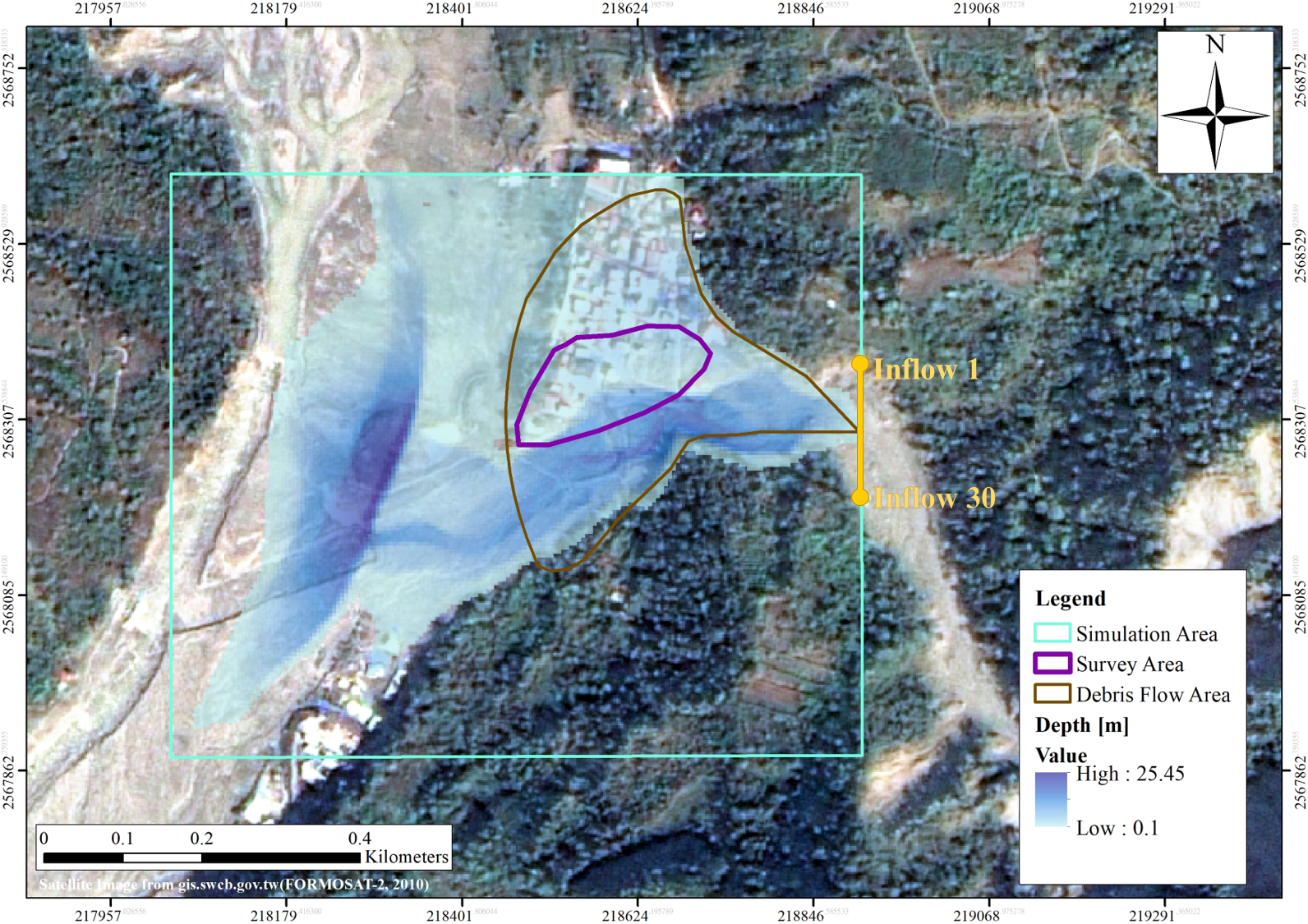}\\
  \hbox{\hspace{2.5cm}c)}
\includegraphics[width=8cm]{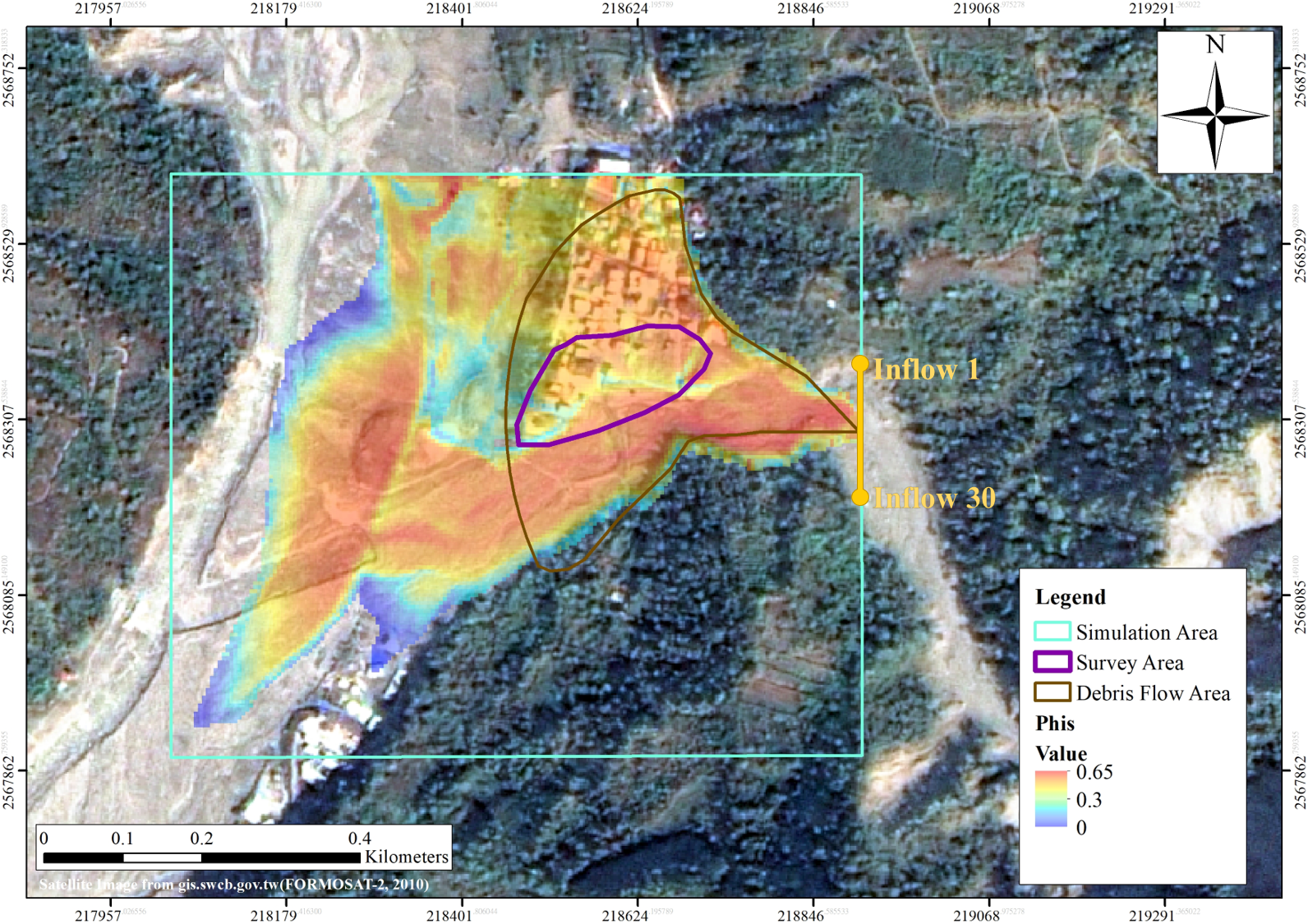}
\caption{Simulation of DF004 event. (a,b) The flow thickness. (c) The solid volume fraction. 
The area enclosed by the cyan lines is the domain in computation.  }
\label{DF004_Fig}
\end{figure}

In the computation domain there are 27,720 meshes (including 3 ghost cells at each 
boundaries), covering an area of $870\times740$ m$^2$. 
The mesh size is $\Delta x =  \Delta y = 5$ m.
The inflow condition consists of the inflow locations, the associate fluxes 
and durations at the boundaries of the computational domain.
In general, the inflow location and fluxes (hydrograph) are estimated
by a 1D computation with empirical relation  (based on the rainfall records and the peak discharge 
with respect to a return period of 200 years),
see \cite{nakatani2008development,liu2013effect,nakatani2016case}.  
In the present study, the inflow condition is based on the computed results
by a rainfall run-off model and scaling the 3 hours rainfall record before the
event into 6 minutes (see the top panels of Fig.~\ref{RainFallCondition} 
and report by \cite{SWCB_Tai_2020}), where the
locations with high fluxes are set to be the inflow points of debris flows, cf. 
the yellow section at the eastern boundary of the computational domain in Fig.~\ref{DF004_Fig}. 
Figure~\ref{RainFallCondition} shows the rainfall records (3 hours before the event)
and the estimated inflow flux at the 30 inflow meshes, which coincide with the bar area
in the post-event picture (see Fig.~\ref{DF004_Fig}b). The scenario period is 12 mins, and 
it takes 202.8 s and 119.43 s for the GPU-2080-Ti and GPU-Tesla-V100 computations, respectively.



Figure~\ref{DF004_Fig} illustrates the flow thickness (panels a and b) and 
the associated distribution of volume fraction for the solid constituent (panel c)
at $t = 12$ mins in the computation.  Here we recall that the areas with flow depth 
less than 10 cm are isolated.
Figure~\ref{DF004_Fig}a is integrated with the satellite image taken before the event,
and panels b and c are integrated with post-event images. Both of the images are 
provided by the Big Geospatial Information System (BigGIS), a web-based database created by
the Soil and Water Conservation Bureau (SWCB), Taiwan.
The brown line marks the plausible debris-flow-endangered
area delineated by the SWCB, Taiwan, 
using the Ikeya's empirical formula~\citep{ikeya1981method}. 
It is found that the flow body approximately coincides with the brown-line-marked area.
In addition, the purple-line-circulated area depicts the survey area post disaster \citep{FastReport2009},
in which  the volume of the deposited debris was estimated up to 70,000 m$^3$ lying in 
the circulated area, while it is about 66,066 m$^3$ in the computation by 
MoSES\_2PDF.

As shown in Fig.~\ref{DF004_Fig}bc, the inflow locations in the computation 
coincide with the bare area in the post-event image,
which approximately mimics the plausible flow paths from upstream during the event.
In the computation, the flow thickness may reach more than 10 cm in the area of village
(see Fig.~\ref{DF004_Fig}ab). Nevertheless, the transparency feature reveals that 
houses and buildings in the northern part of the village was neither flushed away nor buried. 
One of the possible reasons is that the flow thickness is thin, although it is more than 10 cm.
The other reason is the fact that the applied DEM does not include the buildings, 
so that the impacts of the building cannot be reproduced and less mass was collected 
in the purple-line-circulated area.
In Fig.~\ref{DF004_Fig}c, it is found that the variation of the solid concentration
approximately indicates the topography, where the locations of low solid concentration are
in coincidence with the positions of high topography gradient, especially with the 
contour of the river bank. 
The village is mainly covered
by the flow body of high solid concentration, which is speculated to be caused 
by the relatively flat and gentle slope in the village area.




\section{Concluding remarks}
\label{Sect5}
In the present study, we have presented a GPU-accelerated simulation tool, MoSES\_2PDF, for two-phase
grain-fluid shallow flows, where the input and output data can be 
integrated with the GIS system for subsequent application. In comparison
with the conventional CPU computation, the present GPU-CUDA implementation
can enhance the efficiency by up to 230 folds, depending on the PC/workstations, GPU
cards and the mesh numbers in the computation domain.
According to the types of initial condition, there are two modes 
by MoSES\_2PDF: Mode-I is for the landslide
type and Mode-II is for debris flow type. 
Two application examples are illustrated: the 2009 Hsiaolin landslide event 
and the DF004 debris flow event. Each of them
is corresponding to the associated mode, and sound agreements  are delievered
with respect to the post-event satellite images as well as field survey.

The MoSES\_2PDF is able to provide the identical results either with the
RTX-2080-Ti or the Tesla-V100 GPU cards, although the results are is slightly deviated from the ones
computed by the CPU-code. This discrepancy is suspected to be caused by the different structure in 
the source codes and some float/double operation.
In addition, the utilization of MoSES\_2PDF is not limited by the above-listed GPU-cards. 
In our tests, MoSES\_2PDF can well run with many other NVIDIA GPU cards, such as models GTX-1060, 
GTX-1080-Ti, RTX-2070, RTX-3070, RTX-3090. The requirement is that the 
CUDA vers. 9.2 (or higher version) should be installed.

As well, the MoSES\_2PDF is also equipped with three output facilities for illustrating the computed results. 
In addition to the asc-format by GIS and conventional contour maps, the ANSI-Platform provides 
with the animated 3D graphics, where the user can interact with 3D scenes for demonstrating 
or investigating the computed results.  Besides, taking benefits of the high computational efficiency, 
MoSES\_2PDF is able to perform abundant
plausible scenarios in short time. It may help us  to figure out the best fitted or optimal parameters 
by means of back-calculation of historical events for hazard assessment or prediction before the event taking place.
The features and facilities  of MoSES\_2PDF make it suited in an efficient way for engineering applications 
as well as for the hazard assessment or evaluation of plausible disaster mitigation countermeasures.


\paragraph{Codeavailability} The codes for producing the results, the associated illustration facilities
  and the numerical examples  are available on request to th ecorresponding author.
  It is remarked that, following the government regulation, only the DEMs with
resolution of $20\times20$ m can be provided. 

\paragraph{Acknowledgements}
The financial support of the Ministry of Science and Technology, Taiwan (MOST 109-2221-E-006-022-) 
and the Soil and Water Conservation Bureau, Council of Agriculture, Taiwan (SWCB-109-269) are acknowledged.
In addition, the authors thank the Taiwan Computing Cloud (TWCC) for the utilization of 
the GPU computation (NVIDIA Tesla-V-100-SXM2-32GM).





\bibliographystyle{unsrtnat}
\bibliography{arXiv_0a}  

\begin{thebibliography}{63}
\providecommand{\natexlab}[1]{#1}
\providecommand{\url}[1]{\texttt{#1}}
\expandafter\ifx\csname urlstyle\endcsname\relax
  \providecommand{\doi}[1]{doi: #1}\else
  \providecommand{\doi}{doi: \begingroup \urlstyle{rm}\Url}\fi

\bibitem[Pitman and Le(2005)]{pitman2005two}
E~Bruce Pitman and Long Le.
\newblock A two-fluid model for avalanche and debris flows.
\newblock \emph{Philosophical Transactions of the Royal Society A:
  Mathematical, Physical and Engineering Sciences}, 363\penalty0
  (1832):\penalty0 1573--1601, 2005.

\bibitem[Li and Duffy(2011)]{li2011fully}
Shuangcai Li and Christopher~J Duffy.
\newblock Fully coupled approach to modeling shallow water flow, sediment
  transport, and bed evolution in rivers.
\newblock \emph{Water Resources Research}, 47\penalty0 (3), 2011.

\bibitem[Pudasaini(2012)]{pudasaini2012general}
Shiva~P Pudasaini.
\newblock A general two-phase debris flow model.
\newblock \emph{Journal of Geophysical Research: Earth Surface}, 117\penalty0
  (F3), 2012.

\bibitem[Pudasaini and Mergili(2019)]{pudasaini2019multi}
Shiva~P Pudasaini and Martin Mergili.
\newblock A multi-phase mass flow model.
\newblock \emph{Journal of Geophysical Research: Earth Surface}, 124\penalty0
  (12):\penalty0 2920--2942, 2019.

\bibitem[Pudasaini and Hutter(2003)]{pudasaini2003rapid}
Shiva~P Pudasaini and Kolumban Hutter.
\newblock Rapid shear flows of dry granular masses down curved and twisted
  channels.
\newblock \emph{Journal of Fluid Mechanics}, 495:\penalty0 193, 2003.

\bibitem[Bouchut and Westdickenberg(2004)]{Bouchut2004}
F.~Bouchut and M.~Westdickenberg.
\newblock Gravity driven shallow water models for arbitrary topography.
\newblock \emph{Comm. Math. Sci}, 2:\penalty0 359--389, 2004.

\bibitem[Tai and Kuo(2008)]{TaiKuo}
Yih-Chin Tai and Ch-Y Kuo.
\newblock A new model of granular flows over general topography with erosion
  and deposition.
\newblock \emph{Acta Mechanica}, 199\penalty0 (1-4):\penalty0 71--96, 2008.

\bibitem[Tai et~al.(2012)Tai, Kuo, and Hui]{tai2012alternative}
Yih-Chin Tai, Chih-Yu Kuo, and Wai-How Hui.
\newblock An alternative depth-integrated formulation for granular avalanches
  over temporally varying topography with small curvature.
\newblock \emph{Geophysical \& Astrophysical Fluid Dynamics}, 106\penalty0
  (6):\penalty0 596--629, 2012.

\bibitem[Luca et~al.(2016)Luca, Tai, Kuo, et~al.]{luca2016shallow}
Ioana Luca, Yih-Chin Tai, Chih-Yu Kuo, et~al.
\newblock \emph{Shallow geophysical mass flows down arbitrary topography}.
\newblock Springer, 2016.

\bibitem[Rauter and Tukovi{\'c}(2018)]{rauter2018finite}
Matthias Rauter and {\v{Z}}eljko Tukovi{\'c}.
\newblock A finite area scheme for shallow granular flows on three-dimensional
  surfaces.
\newblock \emph{Computers \& Fluids}, 166:\penalty0 184--199, 2018.

\bibitem[Tai et~al.(2021)Tai, Vides, Nkonga, and Kuo]{tai2021multi}
Yih-Chin Tai, Jeaniffer Vides, Boniface Nkonga, and Chih-Yu Kuo.
\newblock Multi-mesh-scale approximation of thin geophysical mass flows on
  complex topographies.
\newblock \emph{Communications in Computational Physics}, 29\penalty0
  (1):\penalty0 148--185, 2021.

\bibitem[Liu and Mei(1989)]{liu1989slow}
Ko~Fei Liu and Chiang~C Mei.
\newblock Slow spreading of a sheet of bingham fluid on an inclined plane.
\newblock \emph{Journal of fluid mechanics}, 207:\penalty0 505--529, 1989.

\bibitem[Coussot(1997)]{coussot1997mudflow}
Philippe Coussot.
\newblock Mudflow rheology and dynamics.
\newblock \emph{Balkema editions, Rotterdam, The Netherlands}, 1997.

\bibitem[Huang and Garcia(1998)]{huang1998herschel}
Xin Huang and Marcelo~H Garcia.
\newblock A herschel--bulkley model for mud flow down a slope.
\newblock \emph{Journal of fluid mechanics}, 374:\penalty0 305--333, 1998.

\bibitem[Ancey(2007)]{ancey2007plasticity}
Christophe Ancey.
\newblock Plasticity and geophysical flows: a review.
\newblock \emph{Journal of Non-Newtonian Fluid Mechanics}, 142\penalty0
  (1-3):\penalty0 4--35, 2007.

\bibitem[Bagnold(1954)]{bagnold1954experiments}
Ralph~Alger Bagnold.
\newblock Experiments on a gravity-free dispersion of large solid spheres in a
  newtonian fluid under shear.
\newblock \emph{Proceedings of the Royal Society of London. Series A.
  Mathematical and Physical Sciences}, 225\penalty0 (1160):\penalty0 49--63,
  1954.

\bibitem[Takahashi(1978)]{takahashi1978mechanical}
Tamotsu Takahashi.
\newblock Mechanical characteristics of debris flow.
\newblock \emph{Journal of the Hydraulics Division}, 104\penalty0 (8):\penalty0
  1153--1169, 1978.

\bibitem[O’Brien and Julien(1985)]{o1985physical}
J.S. O’Brien and P.Y. Julien.
\newblock Physical properties and mechanics of hyperconcentrated sediment
  flows.
\newblock \emph{Proc. ASCE HD Delineation of landslides, flash flood and debris
  flow Hazards}, 1985.

\bibitem[O'Brien and Julien(1988)]{o1988laboratory}
Jim~S. O'Brien and Pierre~Y. Julien.
\newblock Laboratory analysis of mudflow properties.
\newblock \emph{Journal of hydraulic engineering}, 114\penalty0 (8):\penalty0
  877--887, 1988.

\bibitem[O'Brien et~al.(1993)O'Brien, Julien, and Fullerton]{o1993two}
James~S O'Brien, Pierre~Y. Julien, and W.T. Fullerton.
\newblock Two-dimensional water flood and mudflow simulation.
\newblock \emph{Journal of hydraulic engineering}, 119\penalty0 (2):\penalty0
  244--261, 1993.

\bibitem[Iverson(1997)]{iverson1997physics}
Richard~M Iverson.
\newblock The physics of debris flows.
\newblock \emph{Reviews of geophysics}, 35\penalty0 (3):\penalty0 245--296,
  1997.

\bibitem[Iverson and Denlinger(2001)]{iverson2001flow}
Richard~M Iverson and Roger~P Denlinger.
\newblock Flow of variably fluidized granular masses across three-dimensional
  terrain: 1. coulomb mixture theory.
\newblock \emph{Journal of Geophysical Research: Solid Earth}, 106\penalty0
  (B1):\penalty0 537--552, 2001.

\bibitem[Pudasaini et~al.(2005)Pudasaini, Wang, and
  Hutter]{pudasaini2005modelling}
SP~Pudasaini, Y~Wang, and K~Hutter.
\newblock Modelling debris flows down general channels.
\newblock \emph{Natural Hazards and Earth System Sciences}, 5\penalty0
  (6):\penalty0 799--819, 2005.

\bibitem[Tai and Kuo(2012)]{tai2012modelling}
Y.-C. Tai and C.-Y. Kuo.
\newblock Modelling shallow debris flows of the {C}oulomb-mixture type over
  temporally varying topography.
\newblock \emph{Natural Hazards and Earth System Sciences}, 12\penalty0
  (2):\penalty0 269--280, 2012.

\bibitem[Egashira(1997)]{egashira1997constitutive}
Shinji Egashira.
\newblock Constitutive equations of debris flow and their applicability.
\newblock In \emph{1st. International Conference on Debris-Flow Hazards
  Mitigation, ASCE, 1997}, pages 340--349, 1997.

\bibitem[Egashira(2007)]{egashira2007review}
Shinji Egashira.
\newblock Review of research related to sediment disaster mitigation.
\newblock \emph{Journal of Disaster Research}, 2\penalty0 (1):\penalty0 11--18,
  2007.

\bibitem[Nakatani et~al.(2008)Nakatani, Wada, Satofuka, and
  Mizuyama]{nakatani2008development}
Kana Nakatani, Takashi Wada, Yoshifumi Satofuka, and Takahisa Mizuyama.
\newblock Development of “kanako 2d (ver. 2.00),” a user-friendly one-and
  two-dimensional debris flow simulator equipped with a graphical user
  interface.
\newblock \emph{International Journal of Erosion Control Engineering},
  1\penalty0 (2):\penalty0 62--72, 2008.

\bibitem[Liu et~al.(2013)Liu, Nakatani, and Mizuyama]{liu2013effect}
Jinfeng Liu, Kana Nakatani, and Takahisa Mizuyama.
\newblock Effect assessment of debris flow mitigation works based on numerical
  simulation by using {K}anako 2d.
\newblock \emph{Landslides}, 10\penalty0 (2):\penalty0 161--173, 2013.

\bibitem[Nakatani et~al.(2016)Nakatani, Hayami, Satofuka, and
  Mizuyama]{nakatani2016case}
Kana Nakatani, Satoshi Hayami, Yoshifumi Satofuka, and Takahisa Mizuyama.
\newblock Case study of debris flow disaster scenario caused by torrential rain
  on kiyomizu-dera, kyoto, japan-using hyper kanako system.
\newblock \emph{Journal of Mountain Science}, 13\penalty0 (2):\penalty0
  193--202, 2016.

\bibitem[Iverson(2009)]{iverson2009elements}
Richard~M Iverson.
\newblock Elements of an improved model of debris-flow motion.
\newblock In \emph{AIP Conference Proceedings}, volume 1145(1), pages 9--16.
  American Institute of Physics, 2009.

\bibitem[Iverson and George(2014)]{iverson2014depth}
Richard~M Iverson and David~L George.
\newblock A depth-averaged debris-flow model that includes the effects of
  evolving dilatancy. i. physical basis.
\newblock \emph{Proceedings of the Royal Society A: Mathematical, Physical and
  Engineering Sciences}, 470\penalty0 (2170):\penalty0 20130819, 2014.

\bibitem[Anderson and Jackson(1967)]{anderson1967fluid}
T~Bo Anderson and Roy Jackson.
\newblock Fluid mechanical description of fluidized beds. equations of motion.
\newblock \emph{Industrial \& Engineering Chemistry Fundamentals}, 6\penalty0
  (4):\penalty0 527--539, 1967.

\bibitem[Pelanti et~al.(2008)Pelanti, Bouchut, and Mangeney]{pelanti2008roe}
Marica Pelanti, Fran{\c{c}}ois Bouchut, and Anne Mangeney.
\newblock A roe-type scheme for two-phase shallow granular flows over variable
  topography.
\newblock \emph{ESAIM: Mathematical Modelling and Numerical Analysis},
  42\penalty0 (5):\penalty0 851--885, 2008.

\bibitem[Pailha and Pouliquen(2009)]{pailha2009two}
Mickael Pailha and Olivier Pouliquen.
\newblock A two-phase flow description of the initiation of underwater granular
  avalanches.
\newblock \emph{Journal of Fluid Mechanics}, 633:\penalty0 115, 2009.

\bibitem[Luca et~al.(2012)Luca, Kuo, Hutter, and Tai]{luca2012modeling}
I~Luca, CY~Kuo, K~Hutter, and YC~Tai.
\newblock Modeling shallow over-saturated mixtures on arbitrary rigid
  topography.
\newblock \emph{Journal of Mechanics}, 28\penalty0 (3):\penalty0 523--541,
  2012.

\bibitem[Bouchut et~al.(2016)Bouchut, Fern{\'a}ndez~Nieto, Mangeney, and
  Narbona~Reina]{bouchut2016two}
Fran{\c{c}}ois Bouchut, Enrique~Domingo Fern{\'a}ndez~Nieto, Anne Mangeney, and
  Gladys Narbona~Reina.
\newblock A two-phase two-layer model for fluidized granular flows with
  dilatancy effects.
\newblock \emph{Journal of Fluid Mechanics, 801, 166-221.}, 2016.

\bibitem[Meng et~al.(2017)Meng, Wang, Wang, and Fischer]{meng2017modeling}
Xiannan Meng, Yongqi Wang, Chun Wang, and Jan-Thomas Fischer.
\newblock Modeling of unsaturated granular flows by a two-layer approach.
\newblock \emph{Acta Geotechnica}, 12\penalty0 (3):\penalty0 677--701, 2017.

\bibitem[He{\ss} et~al.(2017)He{\ss}, Wang, and
  Hutter]{hess2017thermodynamically}
Julian He{\ss}, Yongqi Wang, and Kolumban Hutter.
\newblock Thermodynamically consistent modeling of granular-fluid mixtures
  incorporating pore pressure evolution and hypoplastic behavior.
\newblock \emph{Continuum mechanics and thermodynamics}, 29\penalty0
  (1):\penalty0 311--343, 2017.

\bibitem[He{\ss} et~al.(2019)He{\ss}, Tai, and Wang]{hess2019debris}
Julian He{\ss}, Yih-Chin Tai, and Yongqi Wang.
\newblock Debris flows with pore pressure and intergranular friction on rugged
  topography.
\newblock \emph{Computers \& Fluids}, 190:\penalty0 139--155, 2019.

\bibitem[Mergili et~al.(2017)Mergili, Fischer, Krenn, and
  Pudasaini]{mergili2017r}
Martin Mergili, Jan-Thomas Fischer, Julia Krenn, and Shiva~P Pudasaini.
\newblock r.avaflow v1, an advanced open-source computational framework for the
  propagation and interaction of two-phase mass flows.
\newblock \emph{Geoscientific Model Development}, 10\penalty0 (2):\penalty0
  553--569, 2017.

\bibitem[Mergili(2014-2020)]{webpageavaflow}
M.~Mergili.
\newblock r.avaflow -- the mass flow simulation tool. r.avaflow 2.3 user
  manual.
\newblock \url{https://www.avaflow.org/manual.php}, 2014-2020.

\bibitem[Mergili et~al.(2020)Mergili, Jaboyedoff, Pullarello, and
  Pudasaini]{mergili2020back}
Martin Mergili, Michel Jaboyedoff, Jos{\'e} Pullarello, and Shiva~P Pudasaini.
\newblock Back calculation of the 2017 piz cengalo--bondo landslide cascade
  with r. avaflow: what we can do and what we can learn.
\newblock \emph{Natural Hazards and Earth System Sciences}, 20\penalty0
  (2):\penalty0 505--520, 2020.

\bibitem[Baggio et~al.(2021)Baggio, Mergili, and
  D'Agostino]{baggio2021advances}
Tommaso Baggio, Martin Mergili, and Vincenzo D'Agostino.
\newblock Advances in the simulation of debris flow erosion: The case study of
  the rio gere (italy) event of the 4th august 2017.
\newblock \emph{Geomorphology}, 381:\penalty0 107664, 2021.

\bibitem[Castro et~al.(2011)Castro, Ortega, De~la Asuncion, Mantas, and
  Gallardo]{castro2011gpu}
Manuel~J Castro, Sergio Ortega, Marc De~la Asuncion, Jos{\'e}~M Mantas, and
  Jos{\'e}~M Gallardo.
\newblock Gpu computing for shallow water flow simulation based on finite
  volume schemes.
\newblock \emph{Comptes Rendus M{\'e}canique}, 339\penalty0 (2-3):\penalty0
  165--184, 2011.

\bibitem[Brodtkorb et~al.(2012)Brodtkorb, S{\ae}tra, and
  Altinakar]{brodtkorb2012efficient}
Andr{\'e}~R Brodtkorb, Martin~L S{\ae}tra, and Mustafa Altinakar.
\newblock Efficient shallow water simulations on gpus: Implementation,
  visualization, verification, and validation.
\newblock \emph{Computers \& Fluids}, 55:\penalty0 1--12, 2012.

\bibitem[de~la Asunci{\'o}n et~al.(2013)de~la Asunci{\'o}n, Castro,
  Fern{\'a}ndez-Nieto, Mantas, Acosta, and Gonz{\'a}lez-Vida]{de2013efficient}
Marc de~la Asunci{\'o}n, Manuel~J Castro, Enrique~Domingo Fern{\'a}ndez-Nieto,
  Jos{\'e}~M Mantas, Sergio~Ortega Acosta, and Jos{\'e}~Manuel
  Gonz{\'a}lez-Vida.
\newblock Efficient gpu implementation of a two waves tvd-waf method for the
  two-dimensional one layer shallow water system on structured meshes.
\newblock \emph{Computers \& Fluids}, 80:\penalty0 441--452, 2013.

\bibitem[Aureli et~al.(2020)Aureli, Prost, Vacondio, Dazzi, and
  Ferrari]{aureli2020gpu}
Francesca Aureli, Federico Prost, Renato Vacondio, Susanna Dazzi, and Alessia
  Ferrari.
\newblock A gpu-accelerated shallow-water scheme for surface runoff
  simulations.
\newblock \emph{Water}, 12\penalty0 (3):\penalty0 637, 2020.

\bibitem[Dazzi et~al.(2019)Dazzi, Vacondio, and Mignosa]{dazzi2019integration}
S~Dazzi, R~Vacondio, and P~Mignosa.
\newblock Integration of a levee breach erosion model in a gpu-accelerated 2d
  shallow water equations code.
\newblock \emph{Water Resources Research}, 55\penalty0 (1):\penalty0 682--702,
  2019.

\bibitem[Wu et~al.(2021)Wu, Kubatko, and Chan]{wu2021high}
Xinhui Wu, Ethan~J Kubatko, and Jesse Chan.
\newblock High-order entropy stable discontinuous galerkin methods for the
  shallow water equations: curved triangular meshes and gpu acceleration.
\newblock \emph{Computers \& Mathematics with Applications}, 82:\penalty0
  179--199, 2021.

\bibitem[Tai et~al.(2019)Tai, He{\ss}, and Wang]{tai2019modeling}
Yih-Chin Tai, Julian He{\ss}, and Yongqi Wang.
\newblock Modeling two-phase debris flows with grain-fluid separation over
  rugged topography: Application to the 2009 hsiaolin event, taiwan.
\newblock \emph{Journal of Geophysical Research: Earth Surface}, 124\penalty0
  (2):\penalty0 305--333, 2019.

\bibitem[Kurganov and Tadmor(2000)]{kurganov2000new}
Alexander Kurganov and Eitan Tadmor.
\newblock New high-resolution central schemes for nonlinear conservation laws
  and convection--diffusion equations.
\newblock \emph{Journal of Computational Physics}, 160\penalty0 (1):\penalty0
  241--282, 2000.

\bibitem[Kurganov et~al.(2007)Kurganov, Petrova, et~al.]{kurganov2007second}
Alexander Kurganov, Guergana Petrova, et~al.
\newblock A second-order well-balanced positivity preserving central-upwind
  scheme for the saint-venant system.
\newblock \emph{Communications in Mathematical Sciences}, 5\penalty0
  (1):\penalty0 133--160, 2007.

\bibitem[Harris et~al.(2007)]{harris2007optimizing}
Mark Harris et~al.
\newblock Optimizing parallel reduction in cuda.
\newblock \emph{Nvidia developer technology}, 2\penalty0 (4):\penalty0 70,
  2007.

\bibitem[Gray et~al.(1999)Gray, Wieland, and Hutter]{Gray1999}
J.M.N.T. Gray, M.~Wieland, and K.~Hutter.
\newblock Gravity-driven free surface flow of granular avalanches over complex
  basal topography.
\newblock \emph{Proc. R. Soc. Lond. A}, 455:\penalty0 1841--1875, 1999.

\bibitem[Kuo et~al.(2009)Kuo, Tai, Bouchut, Mangeney, Pelanti, Chen, and
  Chang]{Kuo2009}
C.Y. Kuo, Y.C. Tai, F.~Bouchut, A.~Mangeney, M.~Pelanti, R.F. Chen, and K.J.
  Chang.
\newblock Simulation of {T}saoling landslide, {T}aiwan, based on {S}aint
  {V}enant equations over general topography.
\newblock \emph{Engineering Geology}, 104\penalty0 (3-4):\penalty0 181 -- 189,
  2009.

\bibitem[de'Michieli Vitturi et~al.(2019)de'Michieli Vitturi, Esposti~Ongaro,
  Lari, and Aravena]{de2019imex_sflow2d}
Mattia de'Michieli Vitturi, Tomaso Esposti~Ongaro, Giacomo Lari, and Alvaro
  Aravena.
\newblock Imex\_sflow2d 1.0: a depth-averaged numerical flow model for
  pyroclastic avalanches.
\newblock \emph{Geoscientific Model Development}, 12\penalty0 (1):\penalty0
  581--595, 2019.

\bibitem[Ryoo et~al.(2007)Ryoo, Rodrigues, Stone, Baghsorkhi, Ueng, and
  Hwu]{ryoo2007program}
Shane Ryoo, Christopher Rodrigues, Sam Stone, Sara Baghsorkhi, Sain-Zee Ueng,
  and Wen-mei~W Hwu.
\newblock Program optimization study on a 128-core gpu.
\newblock In \emph{The First Workshop on General Purpose Processing on Graphics
  Processing Units}, pages 30--39. Citeseer, 2007.

\bibitem[Dong et~al.(2011)Dong, Li, Kuo, Sung, Li, Lee, Chen, and
  Lee]{dong2011}
Jia-Jyun Dong, Yun-Shan Li, Chyh-Yu Kuo, Rui-Tang Sung, Ming-Hsu Li, Chyi-Tyi
  Lee, Chien-Chih Chen, and Wang-Ru Lee.
\newblock The formation and breach of a short-lived landslide dam at {H}siaolin
  village, {T}aiwan—part {I}: {P}ost-event reconstruction of dam geometry.
\newblock \emph{Engineering geology}, 123\penalty0 (1):\penalty0 40--59, 2011.

\bibitem[{Soil Water Conserv. Bureau}(2009)]{FastReport2009}
{Soil Water Conserv. Bureau}.
\newblock \textsl{Disaster Fast Report: 2009 typhoon Morakot -- Namasia
  township 003 (in Chinese)}.
\newblock Technical report, {Soil Water Conserv. Bureau} (SWCB), Taiwan, 2009.

\bibitem[Tai et~al.(2020)Tai, Ko, Li, Wu, Kuo, Chen, and Lin]{tai2020idealized}
Yih-Chin Tai, Chi-Jyun Ko, Kun-Ding Li, Yu-Chen Wu, Chih-Yu Kuo, Rou-Fei Chen,
  and Ching-Weei Lin.
\newblock An idealized landslide failure surface and its impacts on the
  traveling paths.
\newblock \emph{Frontiers in Earth Science}, 8:\penalty0 313, 2020.

\bibitem[Kuo et~al.(2011)Kuo, Tai, Chen, Chang, Siau, Dong, Han, Shimamoto, and
  Lee]{kuo2011landslide}
CY~Kuo, Yih-Chin Tai, CC~Chen, KJ~Chang, AY~Siau, JJ~Dong, RH~Han, T~Shimamoto,
  and CT~Lee.
\newblock The landslide stage of the hsiaolin catastrophe: simulation and
  validation.
\newblock \emph{Journal of Geophysical Research: Earth Surface}, 116\penalty0
  (F4), 2011.

\bibitem[{Soil Water Conserv. Bureau}(2020)]{SWCB_Tai_2020}
{Soil Water Conserv. Bureau}.
\newblock \textsl{GIS integrated web-based system for scenario investigation on
  slope failure and sediment-related disasters (in Chinese)}.
\newblock Technical report, SWCB-109-269, Taiwan, 2020.

\bibitem[Ikeya(1981)]{ikeya1981method}
Hiroshi Ikeya.
\newblock A method of designation for area in danger of debris flow.
\newblock \emph{Erosion and sediment transport in Pacific Rim Steeplands},
  132:\penalty0 578--588, 1981.

\end{thebibliography}

\appendix
\section{Model equations}    
\label{AppendixA}

In MoSES\_2PDF, the employed model equations are adopted 
form  the ones in~\cite{tai2019modeling}. 
With the shallowness assumption for the flow body, the depth-integrated 
dimensionless equations of leading-order read
\begin{equation}
     \displaystyle
     \frac{\partial }{\partial t}\left(J_bh^{s}\right)
     +  \frac{\partial }{\partial \xi}\left(J_bh^{s} v_\xi^{s} \right)
     +  \frac{\partial }{\partial \eta}\left(J_bh^{s} v_\eta^{s}\right) = 0
   \label{MassEq_TopoS0}
\end{equation}
%
%
\begin{equation}
     \displaystyle
     \frac{\partial }{\partial t}\left(J_bh^{f}\right)
     +  \frac{\partial }{\partial \xi}\left(J_bh^{f} v_\xi^{f} \right)
     +  \frac{\partial }{\partial \eta}\left(J_bh^{f} v_\eta^{f}\right) = 0
   \label{MassEq_TopoF0}
\end{equation}
for the mass conservation, 
where $J_b = \det{\boldsymbol\Omega}_b$ is the Jacobian determinant on the
basal surface for coordinate transformation,
$v_{\xi,\eta}^{s,f}$ denote
the tangential components of the depth-averaged velocities in $O_{\xi\eta\zeta}$,
respectively. 
 In (\ref{MassEq_TopoS0}) and (\ref{MassEq_TopoF0}), 
 $h^s=h\phi^s$ and $h^f=h\phi^f$ with $\phi^{s,f}$ 
 the depth-averaged volume  fraction, where
 $h$ stands for the flow thickness measured normal
 to the basal topographic surface.

Regarding the momentum balance equations for the solid and fluid constituents,  the
evolutions of ${\mathbf q}^s=\left(J_bh^sv^s_X, J_bh^sv^s_Y\right)^T$ and
${\mathbf q}^f=\left(J_bh^fv^f_X, J_bh^fv^f_Y\right)^T$
are determined by
  \begin{equation}
   \begin{array}{l}
     \displaystyle
     \frac{\partial {\mathbf q}^s}{\partial t}
     +  \frac{\partial {\mathbf F}^s}{\partial \xi }
     +  \frac{\partial {\mathbf G}^s}{\partial \eta}
     =\displaystyle {\mathbf s}_n^s +{\mathbf s}_f^s + {\mathbf s}_d^s + {\mathbf s}_v^s
   \end{array}
   \label{MomentumEqS_Topo}
 \end{equation}
and
  \begin{equation}
   \begin{array}{l}
     \displaystyle
     \frac{\partial {\mathbf q}^f}{\partial t}
     +  \frac{\partial {\mathbf F}^f}{\partial \xi }
     +  \frac{\partial {\mathbf G}^f}{\partial \eta}
     =\displaystyle {\mathbf s}_n^f +{\mathbf s}_f^f + {\mathbf s}_d^f + {\mathbf s}_{vis}^f\,,
   \end{array}
   \label{MomentumEqF_Topo}
 \end{equation}
respectively, where  ${\mathbf F}^{s,f}$ and ${\mathbf G}^{s,f}$ stand for the fluxes,
\begin{equation}
{\mathbf F}^s =\left(
   \begin{array}{l}
     \displaystyle
      J_bh^sv^s_X v_\xi^s + \epsilon J_b h {A}_{11}\overline{N}^{s}\\[2mm]
      J_bh^sv^s_Y v_\xi^s + \epsilon J_b h {A}_{12}\overline{N}^{s}
   \end{array}
   \right),
\label{FluxEqSxi_Topo}
\end{equation}
\begin{equation}
{\mathbf G}^s =\left(
   \begin{array}{l}
     \displaystyle
      J_bh^sv^s_X v_\eta^s+ \epsilon J_b h {A}_{21}\overline{N}^{s}\\[2mm]
      J_bh^sv^s_Y v_\eta^s+ \epsilon J_b h {A}_{22}\overline{N}^{s}
   \end{array}
   \right),
\label{FluxEqSeta_Topo}
\end{equation}
\begin{equation}
{\mathbf F}^f =\left(
   \begin{array}{l}
     \displaystyle
      J_bh^fv^f_X v_\xi^f + \epsilon J_b h {A}_{11}\overline{p}^{f}\\[2mm]
      J_bh^fv^f_Y v_\xi^f + \epsilon J_b h {A}_{12}\overline{p}^{f}
   \end{array}
   \right),
\label{FluxEqFxi_Topo}
\end{equation}
\begin{equation}
{\mathbf G}^f =\left(
   \begin{array}{l}
     \displaystyle
      J_bh^fv^f_X v_\eta^f+ \epsilon J_b h {A}_{21}\overline{p}^{f}\\[2mm]
      J_bh^fv^f_Y v_\eta^f+ \epsilon J_b h {A}_{22}\overline{p}^{f}
   \end{array}
   \right).
\label{FluxEqFeta_Topo}
\end{equation}
Here, we recall that the unit normal vector ${\boldsymbol{n}}$ at the basal surface 
reads ${\boldsymbol{n}} = (n_X, n_Y, n_Z)^T$  in $O_{XYZ}$, 
and $v_{X,Y}^{s,f}$ are the $X$- and $Y$-components of the phase velocity 
projected on the horizontal plane. In Eqs. (\ref{FluxEqSxi_Topo}) to (\ref{FluxEqFeta_Topo}),
$({A}_{ij})={\boldsymbol\Omega}^{-1}_b$ is the inverse of
 the transformation matrix. $\epsilon$ is the aspect ratio, the ratio of the flow characteristic thickness
 to the characteristic length along the basal surface.
$\overline{N}^{s}=c (1 - \alpha_{\rho} ){h^s}/{2} $ stands for the depth-averaged
 solid pressure, where $c=n_Z$ is the $Z$-component of the unit normal vector of the basal surface
 and $\alpha_{\rho}=\rho^f/\rho^s$ denotes the density ratio of the fluid constituent to the solid one.
 Besides, $\overline{p}^f = c\, {h}/{2} $ stands for the  mean fluid pressure.

The right hand side of (\ref{MomentumEqS_Topo}) consists of four source
terms and they are 
\begin{equation}
{\mathbf s}_n^s =J_b p^s_b \left(
   \begin{array}{l}
      \!n_X\!\!\\[2mm]
      \!n_Y\!\!
   \end{array}\!\!
   \right),\;\;
   {\mathbf s}_d^s =-J_b p^s_b \tan\delta_{b}\left(
   \begin{array}{l}
     \displaystyle
      \frac{v_X^s}{||{\boldsymbol v}^s||}\\[2mm]
     \displaystyle
     \frac{v_Y^s}{||{\boldsymbol v}^s||}
   \end{array}\!\!
   \right),
\label{Src_nd_Topo}
\end{equation}
\begin{equation}
{\mathbf s}_f^s =- \epsilon \alpha_\rho \phi^s \left(\!\!\!
   \begin{array}{l}
     \displaystyle
      {A}_{11}\frac{\partial \left(J_b h \overline{p}^f\right)}{\partial\xi }
                                       +{A}_{21}\frac{\partial\left(J_b h \overline{p}^f\right)}{\partial\eta}\\[2mm]
      \displaystyle
      {A}_{12}\frac{\partial \left(J_b h \overline{p}^f\right)}{\partial\xi }
                                       +{A}_{22}\frac{\partial\left(J_b h \overline{p}^f\right)}{\partial\eta}
   \end{array}
   \!\!\!\right),
\label{Src_f_Topo}
\end{equation}
%
%
\begin{equation}
{\mathbf s}_v^s = J_b \alpha_\rho c_D \phi^s \phi^f h \left(
   \begin{array}{l}
     \displaystyle
      v_X^f- v_X^s \\[2mm]
      v_Y^f- v_Y^s 
   \end{array}
   \right),
\label{Src_v_Topo}
\end{equation}
where $ p^s_b$ means the solid basal 
pressure, $\delta_b$ represents the angle of basal friction and $c_D$ the drag coefficient
for the drag force induced by the relative velocity between the constituents.
In (\ref{Src_nd_Topo}), the solid basal 
pressure reads $ p^s_b = h^s \left( c \left( 1 - \alpha_{\rho} \right)- \epsilon^{\chi}  \kappa^s \right)$,
where $\kappa^s$ means the centripetal acceleration due to the local velocity of the solid constituent
and curvature of the basal surface, and it is given by  \citep[e.g.][]{tai2012alternative,tai2019modeling}
%
\[
   \begin{array}{l}
\displaystyle
\kappa^s = \phantom{+}
   v^s_X \frac{\partial n_X}{\partial \xi } v^s_\xi   + v^s_Y \frac{\partial n_Y}{\partial   \xi } v^s_\xi   
+ v^s_Z \frac{\partial n_Z}{\partial  \xi } v^s_\xi \\ \displaystyle
\phantom{\kappa^s =}
+ v^s_X \frac{\partial n_X}{\partial\eta} v^s_\eta + v^s_Y \frac{\partial n_Y}{\partial \eta} v^s_\eta 
+ v^s_Z \frac{\partial n_Z}{\partial \eta} v^s_\eta\,.
\end{array}
\]

The source terms of (\ref{MomentumEqF_Topo}) read
\begin{equation}
{\mathbf s}_n^f =J_b p^f_b \left(
   \begin{array}{l}
      \!n_X\!\!\\[2mm]
      \!n_Y\!\!
   \end{array}\!\!
   \right),\;
   {\mathbf s}_d^f =-\frac{J_b\phi^f h \vartheta_b^f}{\epsilon N_R} \left(
   \begin{array}{l}
     \displaystyle
      {v_X^f}\\[2mm]
     \displaystyle
      {v_Y^f}
   \end{array}\!\!
   \right)
\label{SrcF_nd_Topo}
\end{equation}
\begin{equation}
{\mathbf s}_f^f = \epsilon  \phi^f \left(
   \begin{array}{l}
     \displaystyle
      {A}_{11}\frac{\partial \left(J_b h \overline{p}^f\right)}{\partial\xi }
                                       +{A}_{21}\frac{\partial\left(J_b h \overline{p}^f\right)}{\partial\eta}\\[2mm]
      \displaystyle
      {A}_{12}\frac{\partial \left(J_b h \overline{p}^f\right)}{\partial\xi }
                                       +{A}_{22}\frac{\partial\left(J_b h \overline{p}^f\right)}{\partial\eta}
   \end{array}
   \right)
\label{SrcF_f_Topo}
\end{equation}
\begin{equation}
{\mathbf s}_v^f = - J_b  c_D \phi^s \phi^f h \left(
   \begin{array}{l}
     \displaystyle
      v_X^f- v_X^s \\[2mm]
      v_Y^f- v_Y^s 
   \end{array}
   \right)
\label{SrcF_v_Topo}
\end{equation}
\begin{equation}
{\mathbf s}_{vis}^f = \frac{\epsilon \phi^f}{N_R}  \left(
   \begin{array}{l}
    s_{vis,X}^f \\[2mm]
    s_{vis,Y}^f
   \end{array}
   \right)
\label{SrcF_vis_Topo}
\end{equation}
with
\[
\begin{array}{l}
s_{vis,X}^f=\displaystyle
      2\frac{\partial}{\partial\xi}\Bigl[J_b h \left(A_{11}\partial_{\xi}{v_\xi^f}
                                                                    +A_{21}\partial_{\eta}{v_\xi^f}\right)\Bigr] \\[2mm]\displaystyle
     + \frac{\partial}{\partial\eta}\Bigl[J_b h \left(A_{12}\partial_{\xi}{v_\xi^f}+A_{22}\partial_{\eta}{v_\xi^f}
                                                                      +A_{11}\partial_{\xi}{v_\eta^f}+A_{21}\partial_{\eta}{v_\eta^f}\right)\Bigr]\\[4mm]
s_{vis,Y}^f=\displaystyle
2\frac{\partial}{\partial \eta}\Big[J_b h\left(A_{12}\partial_{\xi}{v_\eta^f}+A_{22}\partial_{\eta}{v_\eta^f}\right)\Bigr]\\[2mm]\displaystyle
+\frac{\partial}{\partial\xi}\Bigl[J_b h \left(A_{12}\partial_{\xi}{v_\xi^f}+A_{22}\partial_{\eta}{v_\xi^f}
       +A_{11}\partial_{\xi}{v_\eta^f}+A_{21}\partial_{\eta}{v_\eta^f}\right)\Bigr]
\end{array}
\]
for both directions. In (\ref{SrcF_nd_Topo}), $p^f_b$ stands for the fluid basal pressure
and $\vartheta^f_b$ represents the fluid friction coefficient.
The fluid basal pressure is determined by $p^f_b=h^f(c-\epsilon^\chi \kappa^f)$ with
%
\[
\begin{array}{l}
\displaystyle
\kappa^f = \phantom{+}v^f_X \frac{\partial n_X}{\partial\xi } v^f_\xi    + v^f_Y \frac{\partial n_Y}{\partial  \xi } v^f_\xi  
                                  + v^f_Z \frac{\partial n_Z}{\partial\xi } v^f_\xi\\[2mm]\displaystyle\phantom{\kappa^f=}
              + v^f_X \frac{\partial n_X}{\partial\eta} v^f_\eta + v^f_Y \frac{\partial n_Y}{\partial\eta} v^f_\eta 
                                  + v^f_Z \frac{\partial n_Z}{\partial\eta} v^f_\eta,
\end{array}
\]
%
the centripetal acceleration with respect to the fluid velocity and the local curvature of the
topographic surface.







\end{document}